\documentclass[fleqn,usenatbib]{mnras}

\usepackage{newtxtext,newtxmath}

\usepackage[T1]{fontenc}
\usepackage{ae,aecompl}
\usepackage{algorithmic} 

\usepackage{graphicx}	
\usepackage{amsmath}	
\usepackage{amssymb}	
\usepackage[ampersand]{easylist}

\title[Halo around charged meteors]{Proton-induced halo formation in charged meteors}

\author[A. \v{S}ilji\'{c} et al.]{
Ante \v{S}ilji\'{c},$^{1,2}$ 
Frane Luni\'{c},$^{1}$ 
Jurica Tekli\'{c},$^{1}$ 
Dejan Vinkovi\'{c}$^{1,3,4}$\thanks{E-mail: dejan@iszd.hr (DV)}
\\
$^{1}$Faculty of Science, University of Split, Rudjera Bo\v{s}kovi\'{c}a 33, HR-21000 Split, Croatia,\\
$^{2}$Ericsson Nikola Tesla d.d., Polji\v{c}ka cesta 39, HR-21000 Split, Croatia\\
$^{3}$Science and Society Synergy Institute, Bana Josipa Jela\v{c}i\'{c}a 22, HR-40000 \v{C}akovec, Croatia\\
$^{4}$Hipersfera d.o.o., Ilica 36, HR-10000 Zagreb, Croatia
}

\date{Accepted XXX. Received YYY; in original form ZZZ}

\pubyear{2017}

\begin{document}
\label{firstpage}
\pagerange{\pageref{firstpage}--\pageref{lastpage}}
\maketitle

\begin{abstract}
Despite a very long history of meteor science, our understanding of meteor ablation and its shocked plasma physics is still far from satisfactory as we are still missing the microphysics of meteor shock formation and its plasma dynamics. Here we argue that electrons and ions in the meteor plasma above $\sim$100 km altitude undergo spatial separation due to electrons being trapped by gyration in the Earth's magnetic field, while the ions are carried by the meteor as their dynamics is dictated by collisions. This separation process charges the meteor and creates a strong local electric field. We show how acceleration of protons in this field leads to the collisional excitation of ionospheric N$_2$ on the scale of many 100 m. This mechanism explains the puzzling large halo detected around Leonid meteors, while it also fits into the theoretical expectations of several other unexplained meteor related phenomena. We expect our work to lead to more advanced models of meteor-ionosphere interaction, combined with the electrodynamics of meteor trail evolution.
\end{abstract}




\section{Introduction}

Estimates of the daily mass input of interplanetary dust particles into the Earth's atmosphere range from 5 to 300 tonnes per day \citep{REFFPlane}. The ablation of those meteoroid particles, as they enter the atmosphere at hypersonic speeds between $\sim$10 km/s and 72 km/s, is manifested as meteors. 
The physics of meteor-atmosphere interaction is powered by the high level of kinetic energy of meteors, with the light production from collisionally excited atoms as the visually most prominent process. From the public policy point of view, large meteors with enough kinetic energy to penetrate deep into the atmosphere and produce ground damage are of the most concern, as recently seen in the case of Chelyabinsk meteor \citep{REFFPopova}. The Chelyabinsk event stirred up the interest for meteor physics, but the need for better understanding of meteor plasma physics has been already growing in recent years as new meteor related phenomena were discovered. 
Detection of meteors above typical 130 km altitude was explained by thermalization of sputtered particles \citep{REFFVinkovicA, REFFPopovaSputter}, but the theoretical model does not go beyond basic physical principles. Radar observations of such high-altitude meteors indicate the importance of magnetic field and suggest the existence of new types of high-altitude meteors that are possibly a result of some electrodynamic instability triggered by a meteor \citep{REFFGao}. Instrumental detection of anomalous meteor sounds, known as electrophones \citep{REFRAND}, also inspired speculations of a possible electrodynamic coupling between meteors and the ionosphere \citep{REFFZgrablic}.

Another phenomenon, still not explained beyond simple narratives, is large scale (up to $\sim$2 km) jet-like features emanating from high-speed meteors at altitudes above 100 km \citep{REFFLeBlanc}. Fragmentation is a preferred explanation for this and many other unusual meteor light properties, but the cases of high separation speeds of fragments at altitudes too high for fragmentation \citep{REFFStokan} and fast (millisecond) high-amplitude flickering of light curves \citep{REFFSpurny} are beyond this simple explanation. Instead, electrical forces due to accumulated charge in the meteoroid body are sought to explain these phenomena \citep{REFFSpurny,REFFStokan}. This would also agree with the radar observations of oscillating meteor plasma \citep{REFFKero} and the phenomenon of meteors triggering sprites (large-scale electrical discharges high above thunderstorms) by creating kilometer-size plasma irregularities in the ionosphere \citep{REFFQin}. Such a large irregularity has been detected in a form of a kilometer sized halo around a Leonid meteor in 2001 at altitudes between 104 km and 110 km \citep{REFFStenbaekNielsen}. This is a surprising discovery because the size of a meteor plasma ball can be of the order of 1 m, as shown by scattering of radar waves \citep{REFFClose}, since the atmospheric mean free path at 105 km altitude is $\sim$0.33 m. A similar halo effect was detected again in 2002 Leonids \citep{Jenniskens2002}, but not as bright as in the case of 2001 Leonid. The authors in \cite{REFFStenbaekNielsen} considered precursor UV photoionization as a possible source of the halo but dismissed it as inefficient.

All these phenomena indicate that the meteor plasma produces electromagnetic effects on distances larger than what can be reached by thermal collisions of hot meteor plasma. The weakly ionized meteor plasma under hypersonic shock conditions produces a self-induced electric field \citep{REFFFarbar}, but the net charge of the meteor is considered neutral. Hence, we are reviving an old idea of electrically charged meteors that has been so far controversial due to a weak theoretical and observational support - except for micrometeoroids \citep{REFFMendis}, which are not the topic of our study.  In Section \ref{subsec:charging} we give a short overview of existing research on meteor charging, with experiments indicating charge separation at hypersonic speeds. However, the influence of the Earth's magnetic field on the charge dynamics in the meteor head (the immediate surrounding of a meteoroid that includes the shocked region) has not been investigated in any detail. Interestingly enough, in \cite{REFFStenbaekNielsen} the authors speculate that "interaction of the ionized particles with the Earth's magnetic field will cause charge separation that can set up strong local electric fields", which in turn would feed plasma instabilities, but dismiss this idea because ions have a small ($\lesssim$1 m) mean free path at these altitudes. In Section \ref{sec:theorymodel} we show how charge separation still can happen and how accelerated protons can travel long enough distances for the halo to emerge.

\section{History of research on meteor charging mechanisms}
\label{subsec:charging}

Research on charging of objects flying through air (or other gases), such as aircrafts or fast projectiles, has a very long history \citep[see][and references within]{Haseborg1}. In aircrafts this charge can produce various dangerous discharges, while measurements of local electric field disturbances due to accumulated charges on flying projectiles can be used for deducing the projectile velocity and position \citep{Haseborg2}. The research on charging and discharging of supersonic projectiles yielded various results dependent on the projectile properties (size, composition, shape, surface properties, etc.), its speed, and the interaction with the launching platform/gun \citep{Haseborg1}.

Increasing projectile speeds to the hypersonic range, more relevant to the meteor physics, introduces a more challenging physics due to the emergence of plasma flow engulfing the projectile. Laboratory measurements of electric field perturbations by flying projectiles revealed charge separation between the object and its wake, and a formation of a charge precursor in front of the bow shock \citep{SerovYarov91}. Photoionization has been invoked as a possible explanation, with charge separation ahead of the body due to the local electric field produced by the charged body and wake. One of the conclusions of this research was that similar phenomena in meteors could produce large scale ionospheric effects \citep{SerovYarov91}. Improved measurements showed that the magnitude of the charge depends on the Mach number $M$ and does not appear until $M>9$, but the charge polarity depends on various experimental details \citep{PilyuginBaulin,Pilyugin93}. Further experimental research on the electron precursor revealed that it depends on the instability of waves in the plasma and radiation intensity ahead of the bow shock \citep{SerovYarov96}.

While such experiments are indicative of a complicated plasma physics in hypersonic flows, the theory of electrically charged meteors has been so far either over-simplistic - e.g. ignoring details of the meteor plasma physics \citep{REFFDokuchayev} - or erroneous - e.g. misinterpreting the electric field across the overall electrically neutral shock wave as an external long-distance electric field required or detected on the ground, as well as ignoring that the thermal motion by itself cannot create a significant charge excess on a scale much larger that the Debye length because it lacks energy to overcome the electrostatic forces \citep{REFFBeech}. During the preparation of our manuscript a new theory of meteor charging appeared that caries resemblance to our analysis \citep{Kelley}. It argues for huge charge separation electric fields and Hall polarization electric fields along tens of km of meteor path, but it lacks any significant theoretical detail and it is based on various erroneous assumptions. This led the authors to overestimate by many order of magnitude the electric field and the spatial scale of charge separation. Readers interested in more detail can compare their theoretical analysis with our Section \ref{sec:theorymodel}. We are working on a separate publication that will scrutinize in more detail all these meteor charging mechanisms presented in the literature.

A far more promising is a recent theoretical description of induced electric fields in meteor plasma trails by external ionospheric large-scale electric fields \citep{REFFDimant}. Since meteor trails are highly elongated conductors, an expected (albeit not explicitly explored so far) side-effect is a local enhancement of the electric field in the meteor head, where the trail starts, especially in a case of trails oriented more parallel to the magnetic field. The detection of Leonid halo \citep{REFFStenbaekNielsen}, however, happened at the time when conditions were unfavorable for this process as the geomagnetic activity was very low\footnote{http://iswa.ccmc.gsfc.nasa.gov/IswaSystemWebApp/} (Kp$\sim$2) and the shallow East-West meteor direction was oriented more perpendicular to the magnetic and electric field. Nonetheless, it is not excluded that trail currents enhance the electrodynamic effects described in our theory.

\section{Theoretical model of electric field generation}
\label{sec:theorymodel}

Meteors ablate mostly in the dynamo layer of the ionosphere between 75 km and 125 km altitude where the electron collision frequency is lower than the electron cyclotron frequency, but the ion collision frequency is still larger than the ion cyclotron frequency. Ions are therefore largely coupled to the neutral gas while electrons are largely decoupled \citep{REFFSchunk}. When meteors move with an angle to the Earth's magnetic field, electrons in the meteor plasma may stay trapped by the magnetic field while ions keep traveling with the meteor. We argue that this facilitates charge accumulation on meteors. The net charge creates an electric field, which can accelerate charged particles that collide with the atmospheric molecules.

This process is the most efficient in the meteor plasma with the highest speed relative to the magnetic field, which is the region of diffuse shock layer. The formation of diffuse shock is enhanced by the intense meteoroid evaporation, which creates a vapor cloud around the body at altitudes of low atmospheric density where the shock front would not form \citep{REFFPopovaVapor}. The side-lobes of the diffuse bow shock layer are the most susceptible to the leakage of magnetized electrons, since this flow of electrons cannot be stopped by the stagnation zone in front of the meteoroid body \citep{REFFSurzhikov}.

The charge separation can be only partially successful because an electric field builds up along the line of charge separation and drives a neutralization current. The magnetic field is largely unperturbed since the magnetic Reynolds number stays below one (see Appendix \ref{sec:magnetic}; notice how this condition breaks down at altitudes below $\sim$100~km). Hence, electrons are trapped into gyration in meteor plasma and cannot respond collectively to neutralize the charge separation (except along the magnetic field lines) despite their relatively high plasma frequency, high speed and small Debye length. Ions, which are unhindered by the magnetic field, also react to the induced electric field, but their mobility is dictated by collisions within the meteor plasma. This makes them slow (the thermal speed of ions is of the order of $\sim$1~km/s) compared to the charge separation timescale dictated by the much larger meteor speed. 
This also results in electrons having a higher mobility than ions in the meteor current that tries to balance the charges, which reduces the analysis to the Ohm's law for electron motion.

The meteor electron conductivity $\sigma_e^m$ is anisotropic and it depends on the direction of magnetic and electric field. In general terms its magnitude can be written as 
\begin{equation}\label{eq:sigma}
\sigma_e^m=\sigma_0^m \chi
\end{equation}
where $\sigma_0^m$ is the plasma conductivity and $\chi$ is a scaling factor influenced by the direction of fields. We are interested in the conductivity along the electric field lines that drive the neutralization current. In the case when the angle between the meteor flight and the Earth's magnetic field $\overrightarrow{B}$ is close to perpendicular, we could use the Pedersen conductivity within the meteor plasma $\chi=\nu_e^2/(\nu_e^2 + \omega_{ce}^2)$, where $\omega_{ce}=|e|B/m_e=10^7$ Hz is the cyclotron frequency of electrons ($m_e$ is the electron mass) and $\nu_e$ is the electron-neutral collision frequency. However, as we show below, this conductivity is reduced somewhat by the Cowling effect. When $\nu_e > \omega_{ce}$ or the meteor flight is parallel to $\overrightarrow{B}$, the conductivity becomes $\sigma_e^m=\sigma_0^m$ (i.e. $\chi=1$).

The charge conservation gives the balance equation 
\begin{equation}\label{eq:balance}
\sigma_e^m E_m = n_e^m e \text{v}_m
\end{equation}
where $E_m$ is the electric field (in units of V/m), $n_e^m$ is the number density of electrons in meteor plasma (in units of m$^{-3}$) and $e$ is the electron charge. The relative speed between charges during separation is approximately  $\text{v}_m$, which is much larger than ion and neutral thermal speeds. Hence, we use $\sigma_0^m=n_e^m e^2 / m_e \nu_e$ and derive by using equation \ref{eq:balance}:
\begin{equation}\label{eq:Em}
 E_m=\frac{m_e \nu_e \text{v}_m}{e\chi}
\end{equation}

We explore this process at 105 km altitude for Leonid meteors having the velocity of $\text{v}_m=71$ km/s, which yields (in the units of V/m)
\begin{equation}\label{eq:EmLeonid}
 E_m  \sim \left\{\begin{array}{lll}
             4 \times 10^{-7} \nu_e & \text{if}\,\,(\nu_e > \omega_{ce} \,\,\text{and}\,\, \overrightarrow{\text{v}}_m \perp \overrightarrow{B})\,\, \text{or}\,\, \overrightarrow{\text{v}}_m \parallel \overrightarrow{B}\\
             4 \times 10^{-7} \nu_e /\chi & \text{if}\,\, (\nu_e < \omega_{ce} \,\,\text{and}\,\, \overrightarrow{\text{v}}_m \perp \overrightarrow{B})  & 
            \end{array}\right.
\end{equation}
Since in Appendix \ref{sec:magnetic} we show that $\nu_e < \omega_{ce}$ in Leonids at such altitudes, the angle between $\overrightarrow{\text{v}}_m$ and $\overrightarrow{B}$ becomes the critical parameter in deriving the strength of electric field $E_m$. The case $\overrightarrow{\text{v}}_m \parallel \overrightarrow{B}$ would yield small electrical fields and fast charge neutralization, but it looks like the field could reach tens or hundreds of V/m when the meteor flight is close to $\overrightarrow{\text{v}}_m \perp \overrightarrow{B}$. We do not exclude a possibility that the electric field might be also boosted by the internal trail currents induced by the external electric fields \citep{REFFDimant}.

However, in the case of $\overrightarrow{\text{v}}_m \perp \overrightarrow{B}$ there would be an additional strong electric field produced by the Hall current within the meteor plasma perpendicular to $\overrightarrow{\text{E}}_m$  (and $\overrightarrow{\text{v}}_m$) and $\overrightarrow{B}$. Such a field configuration results in an enhanced conductivity along the meteor flight direction (Cowling conductivity) that increases $\chi$ and, therefore, reduces $E_m$. In Appendix \ref{Cowling} we show that the equilibrium between the charging due to meteor flight and enhanced neutralization current due to the Cowling conductivity is reached for 
\begin{equation}\label{eq:chi_equilibirium}
 \chi(L)\sim  \frac{\nu_e}{\omega_{ce}}\cdot \frac{L}{L_0}
\end{equation}
where $L_0$ is the meteor width perpendicular to its direction of flight and $L$ is the distance from the tip of the meteor head along the axis of meteor flight. Notice that at the distance 
\begin{equation}\label{eq:Lmax}
L_{max}=L_0\omega_{ce}/\nu_e
\end{equation}
the conductivity reaches the value $\sigma_e^m\sim \sigma_0^m$ and the neutralization current essentially removes the electric field induced by the charge separation. 

We can combine equations \ref{eq:chi_equilibirium} and \ref{eq:Em} to derive the charge separation electric field at position $L$
\begin{equation}\label{eq:EmCowling}
 E_m(L)=\frac{m_e \omega_{ce} \text{v}_m}{e}\cdot \frac{L_0}{L}
\end{equation}
Now we can show that $L_{max}$ is the effective size of the charge separation region. The timescale $\tau_0$ of charge movement at position $L_{max}$ over distance $L_0$ under conductivity $\sigma_0^m$ can be derived from the plasma drift velocity \citep{Wiesemann}, yielding $\tau_0\sim L_0 m_e \nu_e/e E_m(L_{max})$. Combining this with equations \ref{eq:Lmax} and \ref{eq:EmCowling} gives $\tau_0\sim L_0/\text{v}_m$, which is equal to the time of meteor flight (i.e. there is no enough time to maintain the charge separation any more).  

In the case of our Leonid meteor these values become 
\begin{equation}\label{eq:EmLeonidCowling}
 E_m(L)\sim 4\frac{L_0}{L} \left[\frac{V}{m}\right]
\end{equation}
while the transverse electrostatic field within the Leonid's meteor plasma induced by the Hall current is 
\begin{equation}\label{eq:EHLeonid}
 E_H(L)\sim 4\frac{\omega_{ce}}{\nu_e}\cdot\frac{L_0}{L} \left[\frac{V}{m}\right]
\end{equation}
These equations reveal strong enhancements of the electric field in the very front of the meteor head where electric fields of several hundreds of V/m can be reached. However, it must be noticed that: 
\begin{easylist}
\ListProperties(Hide=3,Hang=true,Style*=-- )
& the charge separation electric field $E_m$ decreases linearly with the distance along the meteor path,
& the entire region of separated charge is connected to the meteor head and it moves with the meteor velocity $\overrightarrow{\text{v}}_m$,
& meteoroid fragmentation within the meteor head might enhance the charge separation because of increased $L_0$,
& the electric field configuration within the meteor head (and consequently the charge distribution) is very complicated and cannot be entirely deduced from such a simple approach.
\end{easylist}

\section{Secondary effects due to charge separation}

Accumulated net charge in the meteor head powers some secondary effects that can be detected from the ground. Here we discuss them in more detail.

\subsection{High speed fragmentation}
\label{sec:fragment}

Meteors often fragment during their flight, which can be usually explained by aerodynamic loading. In case of low brightness meteors at altitudes above $\sim$80 km, the fragmentation could be due to a non-uniform evaporation of the meteoroid body, with relatively low transverse speeds of fragments. Hence, a detection of high transverse speeds, up to $\sim$160 m/s, from low brightness meteors at altitudes of $\sim$100 km was a surprise \citep{REFFStokan}. The authors tested several theoretical explanations and concluded that they are inadequate to explain the fragmentation. It is interesting that one of these rejected models was electrostatic repulsion due to a gain of positive charge, but they rejected it because the estimated net charge was on the order of $10^{11}|e|$ ($e$ is the electron charge). We find below that this is not unrealistic and it fits nicely into our theory.

The expected outer boundary of the meteor charge collection is the region of first collisions between the incoming flow and the diffuse shock material. However, the size of the diffuse shock is not well known and its boundary is smoothly connected with the surrounding ionosphere. Scattering of radar waves from the meteor head plasma yielded estimates of the head radius of $r_0\sim 0.5-1$ m \citep{REFFClose,REFFMarshall}. For simplicity we will use $r_0=1$ m radius. If we use the basic definition of the electric field $E_m=|e| N_p^0/4\pi \varepsilon_0 r_0^2$ then the net number of positive charge is $N_p^0=7\times 10^8E_m$. When $E_m$ is on the level of hundreds of V/m then the accumulated charge $N_p^0$ reaches the critical values predicted for the anomalously high speeds of transverse motion of fragmenting faint meteors. Since the total number of electrons in meteor head plasma at 105 km is $>10^{14}$ \citep{REFFClose}, we see that just a tiny fraction of the electrons in the meteor head needs to be extracted to create strong electric fields.

\subsection{Meteor head charge oscillation}
\label{sec:oscillation}

Meteor fragmentation is monitored lately using high-power, large-aperture VHF/UHF radar systems that can detect the meteor head-echo - the radar reflection from the plasma surrounding the meteoroid in the meteor head. Fragmentation creates radar signal pulsations due to interference patterns as fragments drift away from each other. This drift can be calculated when observed simultaneously from several radar locations, but a surprise detection was when pulsation signals were simultaneous at all radar stations \citep{REFFKero}. The simplest explanation is an oscillating size of the meteor head plasma. The same phenomenon is implied from the high time resolution light curves of fireballs that exhibit very short, millisecond flares \citep{REFFSpurny}, which cannot be explained by classical meteor hypersonic flight physics. Instead, the authors evoke meteor charging as a possible explanation, which would lead to periodic discharges that could evaporate a small inner portion of the meteoroid. They speculate that this process can release small fragments into the head plasma, where they evaporate and boost the meteor light production.

In our model the size of the plasma cloud is naturally prone to oscillations. This is a consequence of two balancing processes: one is charge creation and the other is the loss of net charge as the charge moves into the meteor wake or gets ejected by the electrostatic force. The charge production is dominated by the meteor speed and atmosphere density, which do not depend on the meteor net charge, but the charge loss depends on the strength of electrostatic force, which in turn is proportional to the amount of net charge. This creates a feedback loop that powers the charge density oscillations.

The net charge in meteor head is created in collisions of meteor head particles with the incoming atmosphere  (source function $S(t)$) and lost when transported into the meteor wake or get ejected into the ionosphere by the electrostatic force (sink function $L(t)$). We can approximate $S(t)$ as a constant within a narrow range of altitudes because the number of collisions is mainly dominated by the meteor speed and atmospheric density. The sink $L(t)$ is more complicated because the loss of net charge is dictated by the electrostatic force proportional to the amount of accumulated net charge. Hence, $L(t)$ depends on the net charge number density $n_p(t)$ within the meteor head of radius $r_0$ and on the speed of the leaking charge at the meteor head boundary $\text{v}_p(r_0,t)$. We approximate this as $L(t)=\alpha n_p(t)\text{v}_p(t)$, where $\alpha$ is a constant for a given $r_0$.

In steady state $S=L(t)$, with the charge density and escape velocity at their equilibrium $n_p^0$ and $\text{v}_p^0$. Perturbations from the steady state are a consequence of the electric field force removing the positive net charge with an unsteady speed. From the general definition of the force it follows $E_m(r_0,t)\propto d\text{v}_p(r_0,t)/dt$, while the definition of electric force gives $E_m(r_0,t)\propto n_p(t)/r_0^p$ (we expect charge screening in plasma, which makes $p<2$). This yields the escape velocity variations 
\begin{equation}\label{eq:dvdt}
d\text{v}_p(t)/dt=\beta n_p(t)
\end{equation}
where $\beta$ is a constant for a given $r_0$.
The charge conservation dictates the balance equation $dn_p(t)/dt=S-L(t)$. If we consider perturbations from the equilibrium speed $\text{v}_p(t)=\text{v}_p^0 + \delta \text{v}(t)$ and from the equilibrium density $n_p(t)=n_p^0+\delta n_p(t)$ then the charge conservation yields
\begin{equation}\label{eq:chcons}
\frac{d \delta n_p(t)}{dt}= - \alpha n_p^0 \delta \text{v}(t) - \alpha \text{v}_p^0 \delta n_p(t)
\end{equation}
The same perturbation approach can be applied on the escape velocity variations in equation \ref{eq:dvdt} and obtain
\begin{equation}\label{eq:chvelc}
\frac{d \delta \text{v}_p(t)}{dt}= \beta \delta n_p(t)
\end{equation}
After combining equation \ref{eq:chvelc} with the first derivative of equation \ref{eq:chcons}, we obtain
\begin{equation}
\frac{d^2\delta n_p(t)}{dt^2} + \nu \frac{d \delta n_p(t)}{dt} + \omega_0^2  \delta n_p(t) = 0
\end{equation}
where $\nu=\alpha \text{v}_p^0$ and $\omega_0^2=\alpha\beta n_p^0$.

This is the classical Damped Harmonic Oscillator equation, with the solution
\begin{equation}
\delta n_p(t) \propto e^{-\nu t/2}\cos(\omega_1 t + \phi)
\end{equation}
where $\omega_1^2=\omega_0^2 -\nu^2/4$. The quantitative details are not reachable with this level of approximations, but qualitatively we see that the damping depends on the charge sink rate, while oscillations have a higher frequency for larger electric forces.

\subsection{Ionospheric halo around a charged meteor head}
\label{sec:halotheory}

The net charge accumulated in the meteor head would also produce an electric field around the meteor. The surrounding weakly ionized ionosphere reacts to the electric field by driving electric currents. Strong enough currents might induce collisionally excited emission from the neutral atmospheric species, which would explain a large glowing halo detected around a Leonid meteor \citep{REFFStenbaekNielsen}. However, we find this process unlikely since these currents are dominantly due to ionospheric electrons, as they are more mobile than ions, but the electrons are magnetized and accelerate along the magnetic field lines, which prevents them from creating a halo-like structure,

A strong electric field would also eject ions out of the meteor head. Ions are heavy compared to electrons and cannot reach high speeds under these conditions, except in the case of protons. Protons change into hydrogen atoms and back through the collisional process of charge exchange and electron stripping. The presence of hydrogen in bright meteors, including Leonids, has been revealed through hydrogen emission lines \citep{REFFJenniskensC,REFKasuga}. Hydrogen is a product of the meteoroid ablation and it belongs to the "hot" component of a meteor spectrum, which seems to be related to the impact excitation during non-thermal collisions with the atmosphere. This implies that hydrogen is excited within the diffuse shock and, in our case, in the meteor halo. This also agrees with the observation that the intensity of the high-excitation hydrogen emission follows atmospheric nitrogen, not the meteor afterglow \citep{REFFJenniskensC}.

The first positive band of N$_2$ (B$^3\Pi_g$-A$^3\Sigma_u^+$) is the most probable source of the meteor halo emission caused by proton/H collisions with the atmospheric N$_2$: i) it is regularly detected in meteor spectra \citep{REFBorovicka99,REFFJenniskensC,REFKasuga,REFFJehin}, ii) its wavelength range fits the spectral range of the camera ($\sim$700$\pm$200$\mu$m) that imaged the halo \citep{REFFStenbaekNielsen,REFForrest}, and iii) it has short enough radiative lifetimes  \citep{REFForrest} to emit within the high-speed camera frame length of 1 ms (otherwise the imaged halo would be elongated). The plasma temperature responsible for N$_2$ emission is not easily constrained - derived temperatures vary from $\sim$4500~K \citep{REFFJenniskensB} to $\sim$8000~K \citep{REFKasuga}. The reason might be in photon emission from two distinct plasma sources - one is the regular thermalized meteor plasma, while the other is non-equilibrium excitations of N$_2$ molecules during fast one-directional collisions in the diffuse shock, and possibly in meteor halo. Various ratios of equilibrium to non-equilibrium emission would lead to different temperature estimates, which can manifest itself as a bi-modal temperature of meteor plasma with the main spectrum indicating $\sim$4000~K and the second spectrum, probably originating from the shock, showing $\sim$10 000~K \citep{REFBorovicka94}. 

Using these assumptions, we performed a numerical simulation of proton/H transport through the ionosphere under the influence of meteor electric field. The simulation setup and results are presented in the next Section. 

\begin{figure}
  \centering
  \includegraphics[width=0.46\textwidth]{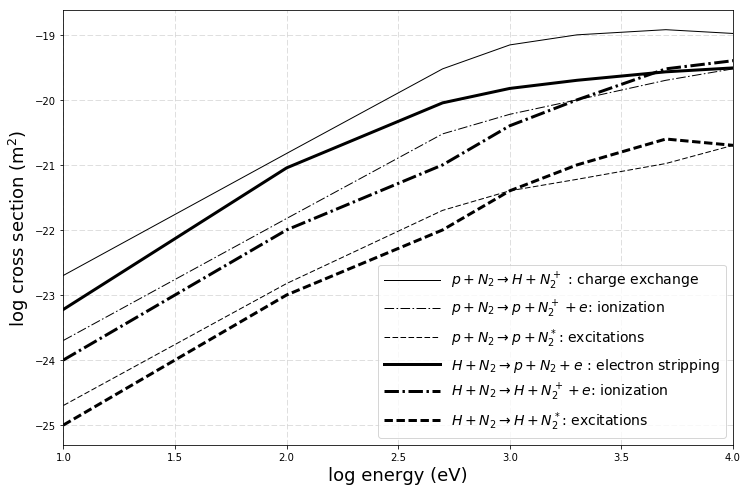}
  \includegraphics[width=0.46\textwidth]{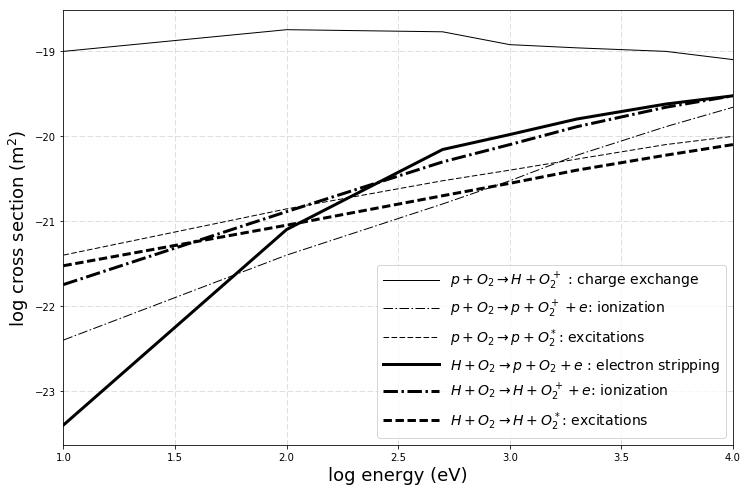}
  \includegraphics[width=0.46\textwidth]{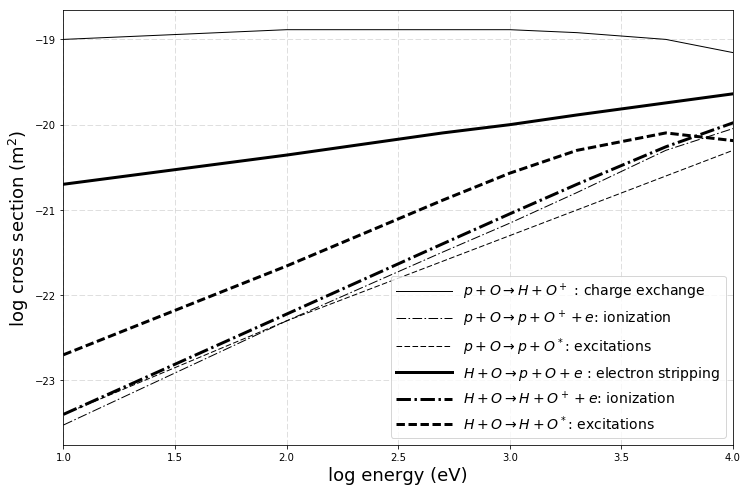}
  \caption{Collisional cross section of proton/hydrogen on N$_2$, O$_2$ and O for ionization, excitations, charge exchange and electron stripping processes.}
  \label{fig:cross_sec}
\end{figure}

\begin{figure}
  \centering
  \includegraphics[width=0.49\textwidth]{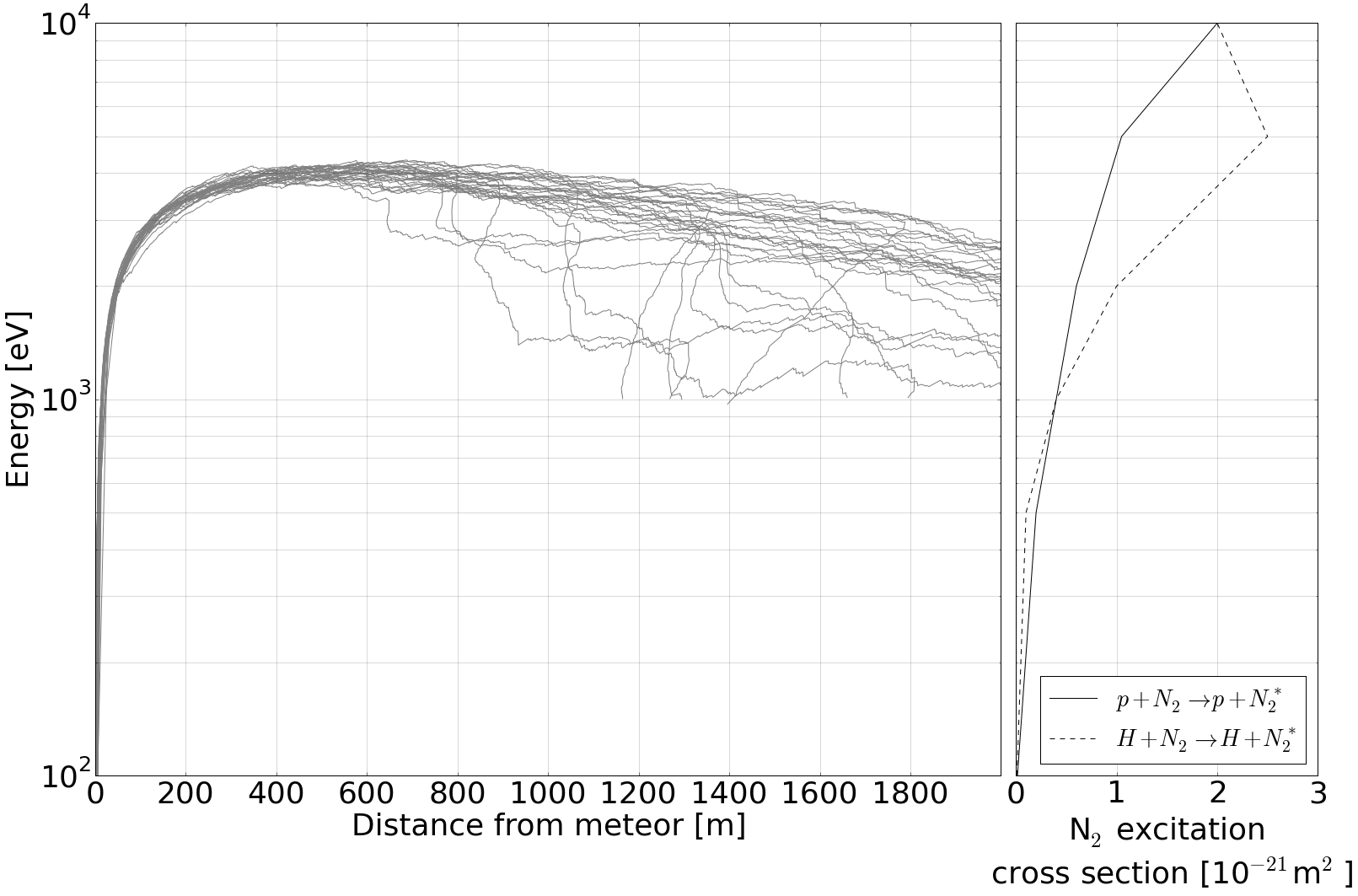}
  \caption{The energy of 30 randomly selected ejected particles relative to their distance from the meteor (left panel). The particles are followed until their energy drops below 1keV or their distance from the meteor exceeds 10 km. The right panel shows the energy dependence of the nitrogen excitation cross section responsible for the light production in our model.}
  \label{fig:energy}
\end{figure}

\begin{figure}
  \centering
  \includegraphics[width=0.49\textwidth]{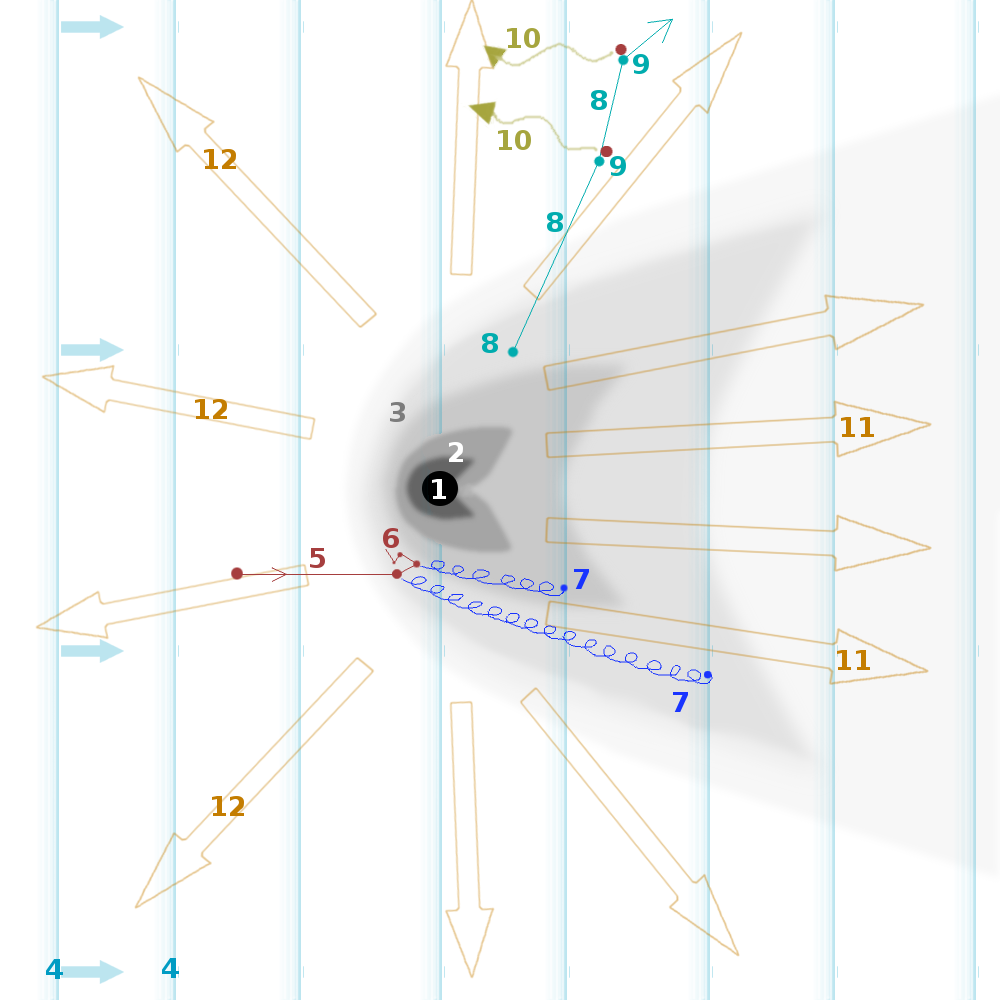}
  \caption{A simplified sketch of the processes used in our meteor charging model, shown from the reference point of a flying meteoroid [1]. The immediate vicinity of the meteoroid has the highest plasma density [2], while the diffuse shock region [3] has a lower density that smoothly connects to the surrounding ionosphere. The magnetic field [4] moves toward the meteor with the meteor speed, together with the atmospheric particles [5] that hit the meteor plasma [6]. The dynamics of ions created in such collisions is dictated by thermal collisions and ions stay coupled to the local meteor plasma [6]. Electrons from these collisions [7] have their collision frequency smaller than the electron cyclotron frequency, which means they start to gyrate and get trapped by the moving magnetic field. This creates a charge separation between ions and electrons, which results in an electric field within the meteor (and its wake) [11] and an electric field around the meteor head [12]. Electron gyration is not entirely free - electrons still experience some collisions. This enables the electric field [11] to reach its maximum when the electron drift reaches its steady state and the further charge separation is stopped. Protons that appear in the meteor plasma [8] (mostly as a product of meteoroid ablation) get accelerated by the electric field and travel long distances [8], slowed down by collisions with the atmospheric species [9]. We are interested in collisions that excite N$_2$ molecules and create photons [10] detectable by the camera system used in the Leonid halo detection \citep{REFFStenbaekNielsen}.}
  \label{fig:sketch}
\end{figure}

\section{Meteor halo modeling}
\label{sec:modeling}

The goal of this model is to reconstruct the brightness profile of the imaged halo around a Leonid meteor \citep{REFFStenbaekNielsen} at 105 km altitude. The simulation is based on proton aurora models \citep{REFFSolomon} tracing ionization, excitations, charge exchange and electron stripping processes on N$_2$, O$_2$ and O. A simulation particle is traced until its energy drops below a threshold value or its distance from the meteoroid becomes too large. 

\subsection{Simulation setup}
\label{subsec:setup}

The environment in our simulation is set to the conditions defined by the detected Leonid halo \citep{REFFStenbaekNielsen} at 105 km altitude. The meteor travels with the speed of 71 km/s from the direction of Leonids radiant positioned at the local azimuth of 86.5$^\circ$ (East from North) and elevation of 22$^\circ$. The camera view direction is 76$^\circ$ (East from North) and 54$^\circ$ elevation. The International Geomagnetic Reference Field model \citep{REFFigrf} gives the local magnetic field strength of 56~963 nT oriented 24.2$^\circ$ (East from North) and 77.4$^\circ$ Down (the magnetic field components are 12~454 nT horizontal, 11~360 nT North, 5~103 nT East and 55~585 nT Down).

The number densities of atmospheric species at 105 km altitude are taken from the MSISE model of the neutral atmosphere \citep{REFFHedin}: atomic oxygen $n_{atm}(O)=3.25\times 10^{17}$ m$^{-3}$, molecular oxygen $n_{atm}(O_2)=7.26\times 10^{17}$ m$^{-3}$ and molecular nitrogen $n_{atm}(N_2)=3.3\times 10^{18}$ m$^{-3}$. The angle between the magnetic field vector and the meteor line of flight was $\sim$74$^\circ$, which is very close to perpendicular and favorable to our theory of the accumulation of net charge. We do not know details of the net charge distribution and the geometry of its electric field, thus we use a simple approximation of a point source. Modeled protons are ejected at distances $r_0=1$ m from the meteor center (meteoroid position). The radial strength of electric field at distance $r$ from the meteoroid is $E(r)=E_0/(r/r_0)^p$, where $E_0$ is the field at $r_0$ and $p$ is a general form of the power law. We expect deviation from $p=2$ because of charge screening in the ionosphere. We can only use an approximate derivation of the screening process (see Appendix \ref{sec:screening}), which yields $2\geq p\geq2/3$. Both $E_0$ and $p$ dictate the electric field range of considerable proton acceleration and we can treat them as free parameters. For the canonical model presented here we use $E_0=$300 V/m and $p=$2/3.

Collisional cross section of proton/hydrogen on N$_2$, O$_2$ and O for ionization, excitations, charge exchange and electron stripping processes are taken from \citep{REFFBasu87} and \citep{REFFRees} and shown in Fig. \ref{fig:cross_sec}. Elastic collisions of proton/hydrogen atoms display highly forward scattered angles, hence we approximate it with a pure forward scattering and, therefore, do not have to simulate this process. Inelastic collisions result in energy loss for proton/hydrogen that depends on the collision energy. Average values for these energy losses are taken from Table 1 in \cite{REFFBasu93}. We use the screened Rutherford phase function to trace proton/hydrogen atoms after collisions. The scattering azimuthal angle $\vartheta$ is obtained from the probability function
\begin{equation}
 \cos\vartheta =1+2\epsilon - \frac{2\epsilon(1+\epsilon)}{1+\epsilon-\Re}
\end{equation}
where $\Re\in (0,1]$ is a random number and
\begin{equation}\label{eq:scattering}
 \epsilon=\epsilon_0 \frac{1000eV}{E}
\end{equation}
is a scattering parameter with $\epsilon_0=0.001$ and kinetic energy $E$ (in eV) of the proton/hydrogen.

A million protons are ejected at distance $d_0=1$ m from the meteoroid along the initial directions $\widehat{n}_p$. The initial direction is selected relative to the meteor direction, with a random azimuthal angle $\phi_{eject}$ between 0 and $360^\circ$ and a random polar angle $\theta_{eject}$ following the probability function 
\begin{equation}\label{eq:ejectangle}
\cos(\theta_{eject})=\cos\vartheta_0-(\cos\vartheta_0-\cos\vartheta_1)\sqrt{\Re},
\end{equation} 
where $\Re$ is a random number and $\vartheta_0=60^\circ$ and $\vartheta_1=150^\circ$. We settled with these angles after testing different choices to fit the observed brightness profile. Some examples of model images produced with different choices of ejection polar angles are shown in Appendix \ref{sec:angles}.

Within the camera exposure time of $T_{exp}=0.001$s the meteor moves with velocity $\overrightarrow{\text{v}}_m$ away from the position it had at the start of the exposure, which gives the initial proton ejection position $\overrightarrow{r}_p=\Re\,T_{exp}\overrightarrow{\text{v}}_m + d_0\widehat{n}_p$, where $\Re\in (0,1]$ is a random number. We set the initial proton velocity equal to $\overrightarrow{\text{v}}_m$. The proton is then accelerated by the electric field $E(r)$. The proton switches to hydrogen during charge exchange collisions and hydrogen turns to proton during electron stripping collisions.

\begin{figure}
  \centering
  \includegraphics[width=0.49\textwidth]{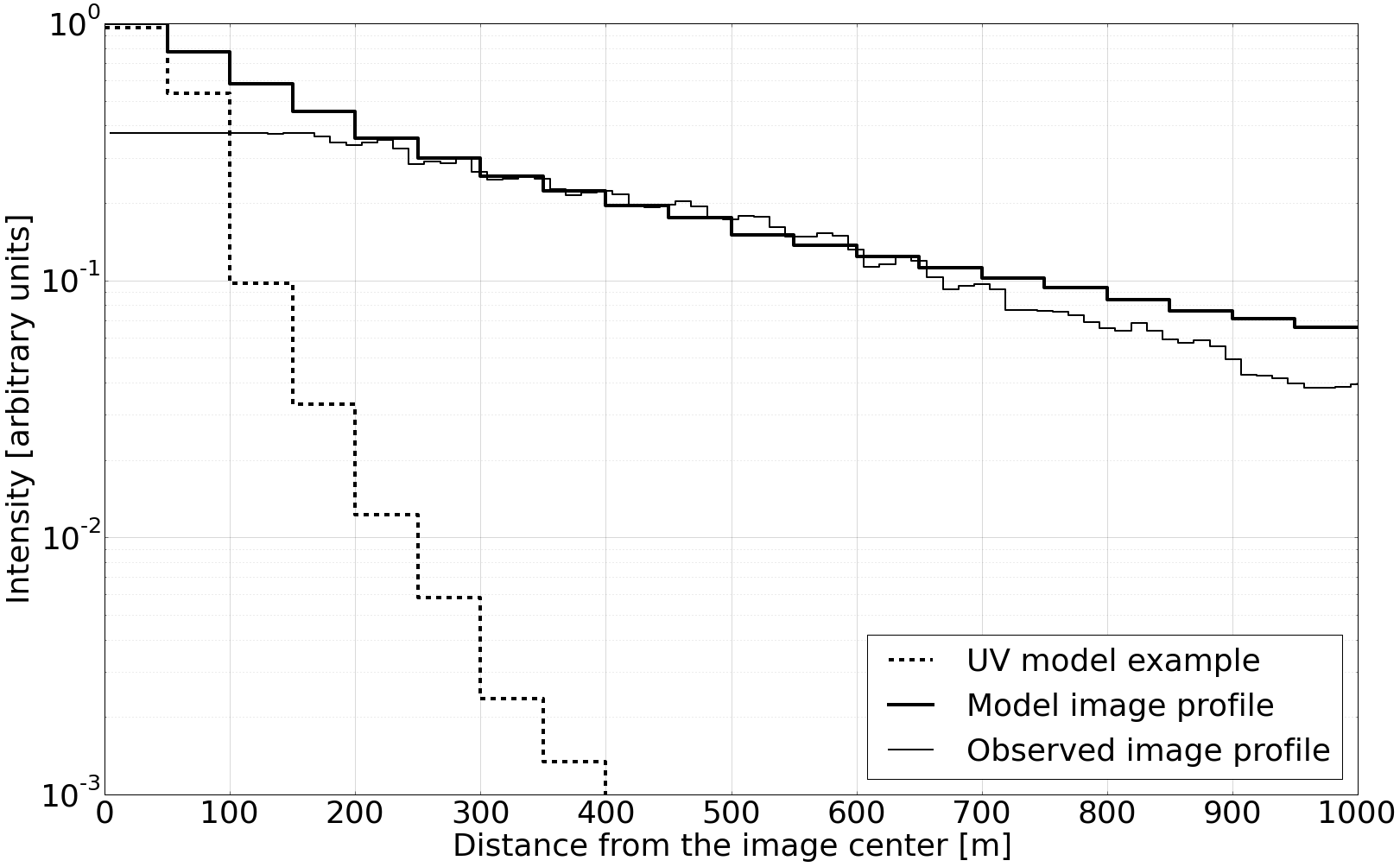}
  \caption{The intensity profile of the halo model image from the view of the camera (thick solid line). The cut off is taken from the image center along the direction perpendicular to the meteor flight. We show how it compares with the observed image profile (thin solid line) taken from Fig. \ref{Fig:mainFig} in \citet{REFFStenbaekNielsen} (from the frame at 104.8 km altitude). The flat part of the observed intensity profile at distances $\lesssim$200 m is due to pixel saturation and comparison with the model is not possible in that region. Brightness oscillations on the scale of $\sim$50 m are artifacts of pixel size. For comparison we also show a typical intensity profile for a UV model of meteor halo with immediate photon emission (dotted line), which emphasizes the inability of precursor UV radiation to explain the size of the halo. }
  \label{fig:canonical_intensity}
\end{figure}

\begin{figure}
  \centering
  \includegraphics[width=0.49\textwidth]{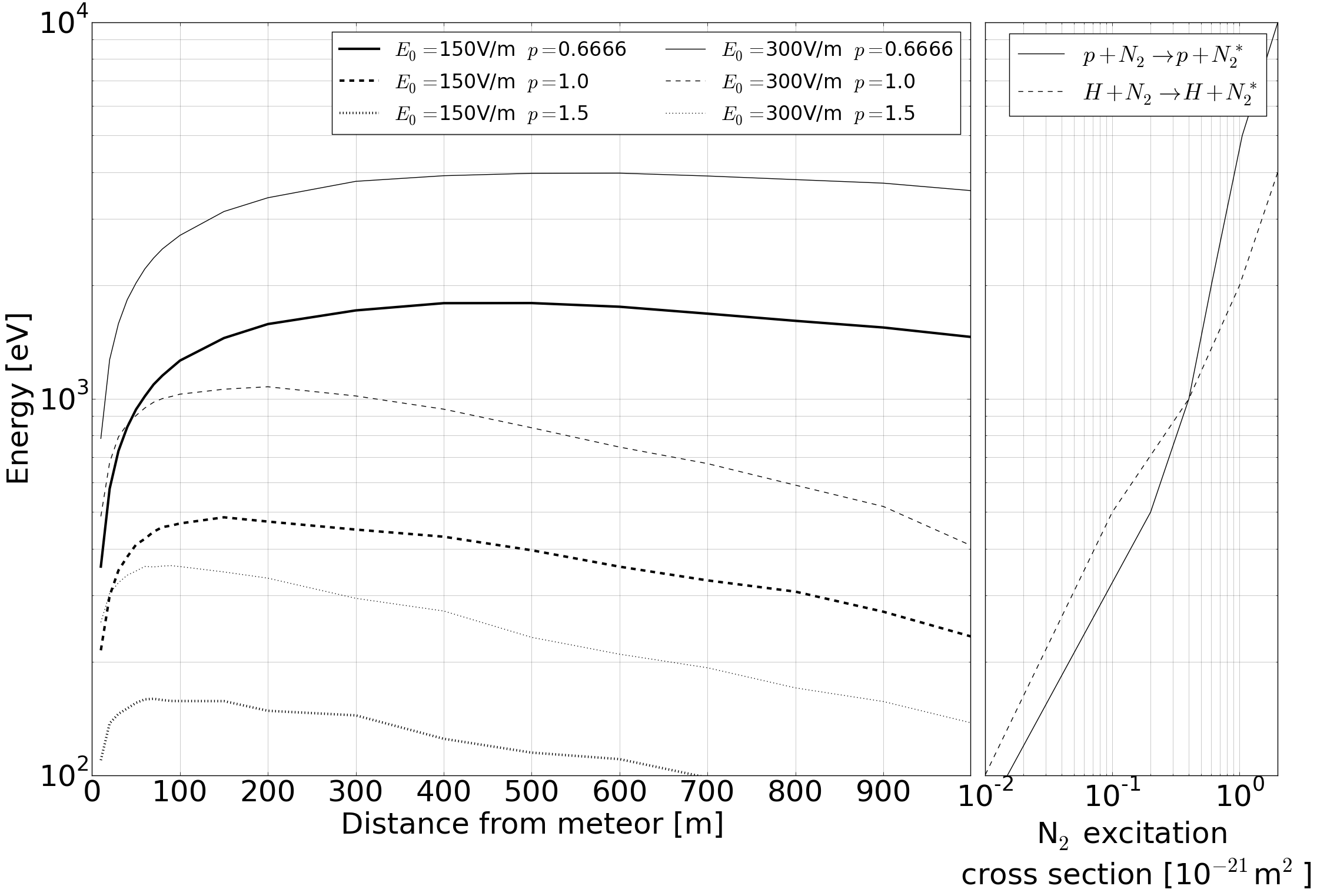}
  \caption{The energy of particles ejected under different electric fields (left panel). Each line is a median of energies of 50 randomly ejected particles accelerated by the electric field $E(r)=E_0/(r/r_0)^p$, where $r$ is the distance from the meteoroid and $r_0=1$ m. The right panel shows the energy dependence of the nitrogen excitation cross section responsible for the light production in our model.}
  \label{fig:change_p_E0}
\end{figure}

\begin{figure}
  \centering
  \includegraphics[width=0.46\textwidth]{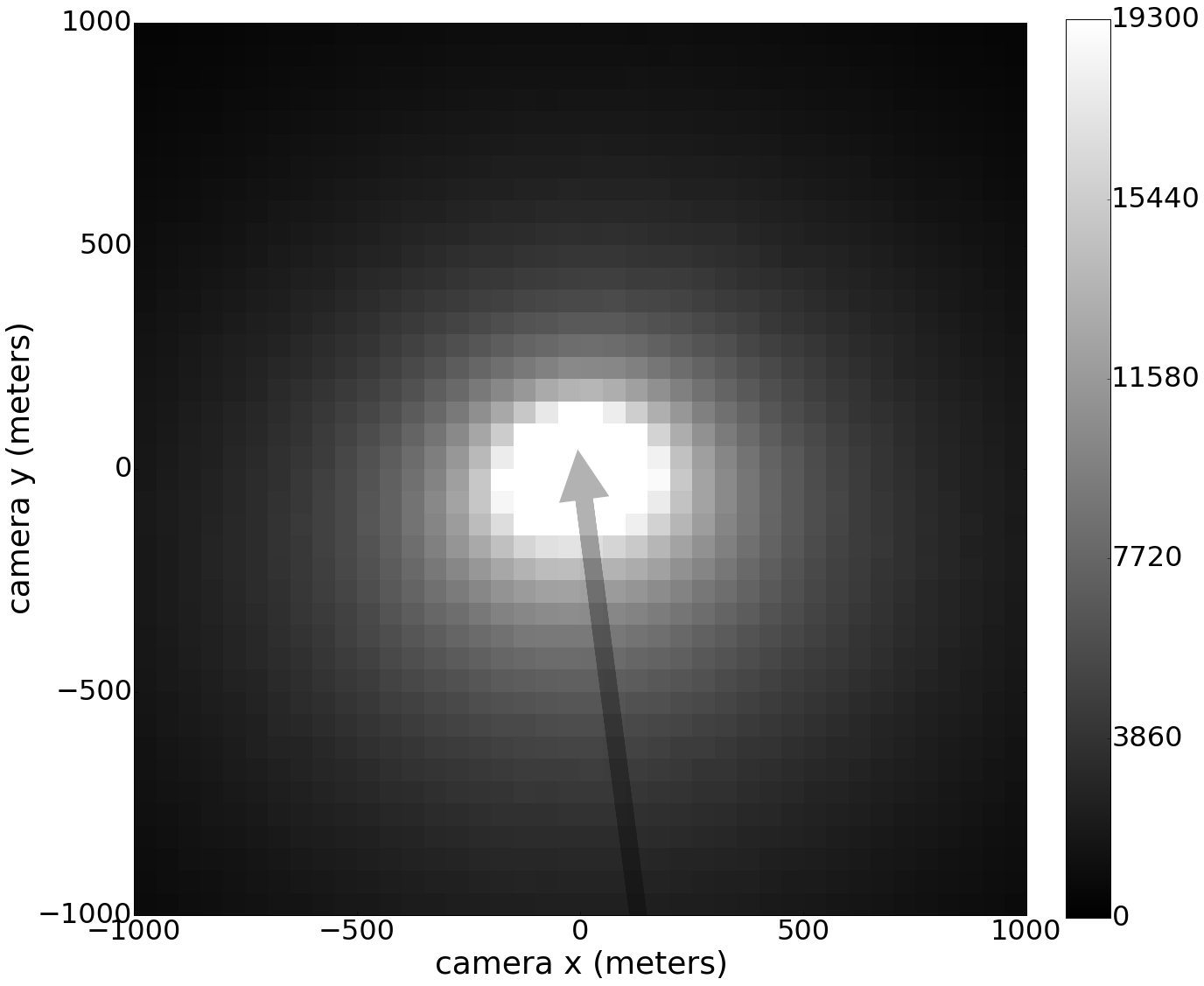}
  \includegraphics[width=0.46\textwidth]{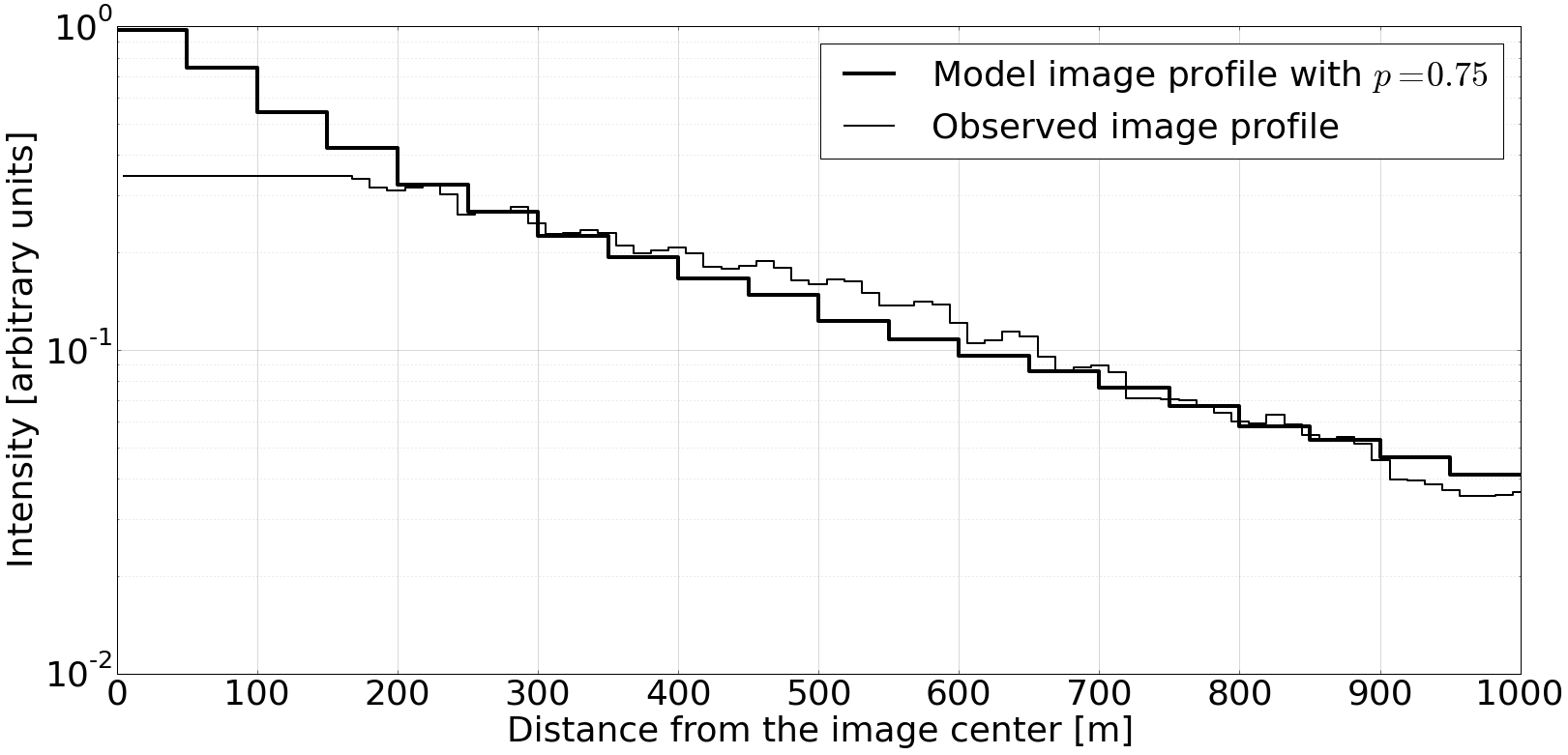}
  \caption{The simulated halo image and the intensity profile using a model where the electric field follows the $p=0.75$ power law. An improvement is visible in matching the modeled and measured brightness profiles compared to the canonical model in Fig. \ref{fig:canonical_intensity}. }
  \label{fig:p=075}
\end{figure}

\begin{figure}
  \centering
  \includegraphics[width=0.46\textwidth]{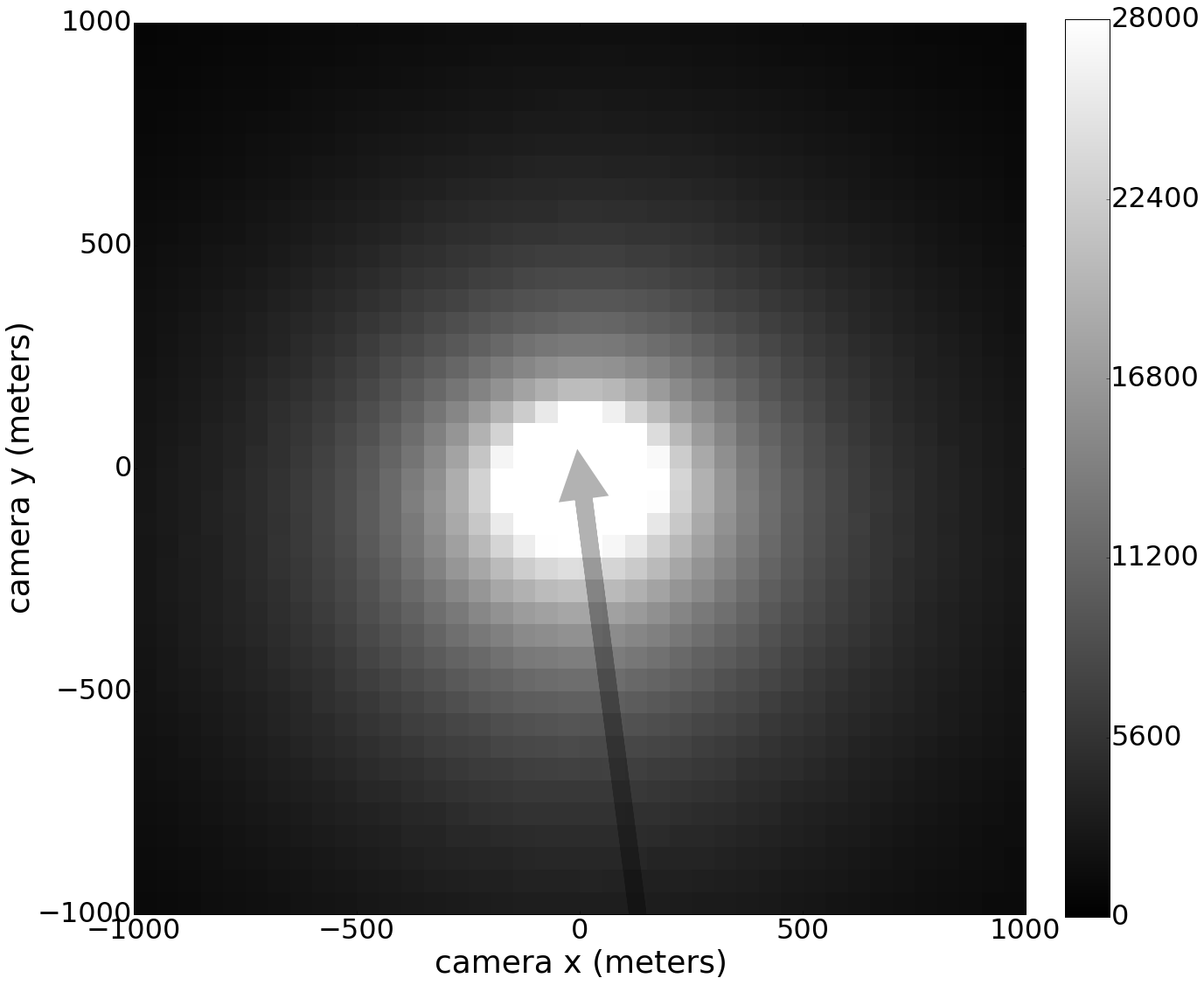}
  \includegraphics[width=0.46\textwidth]{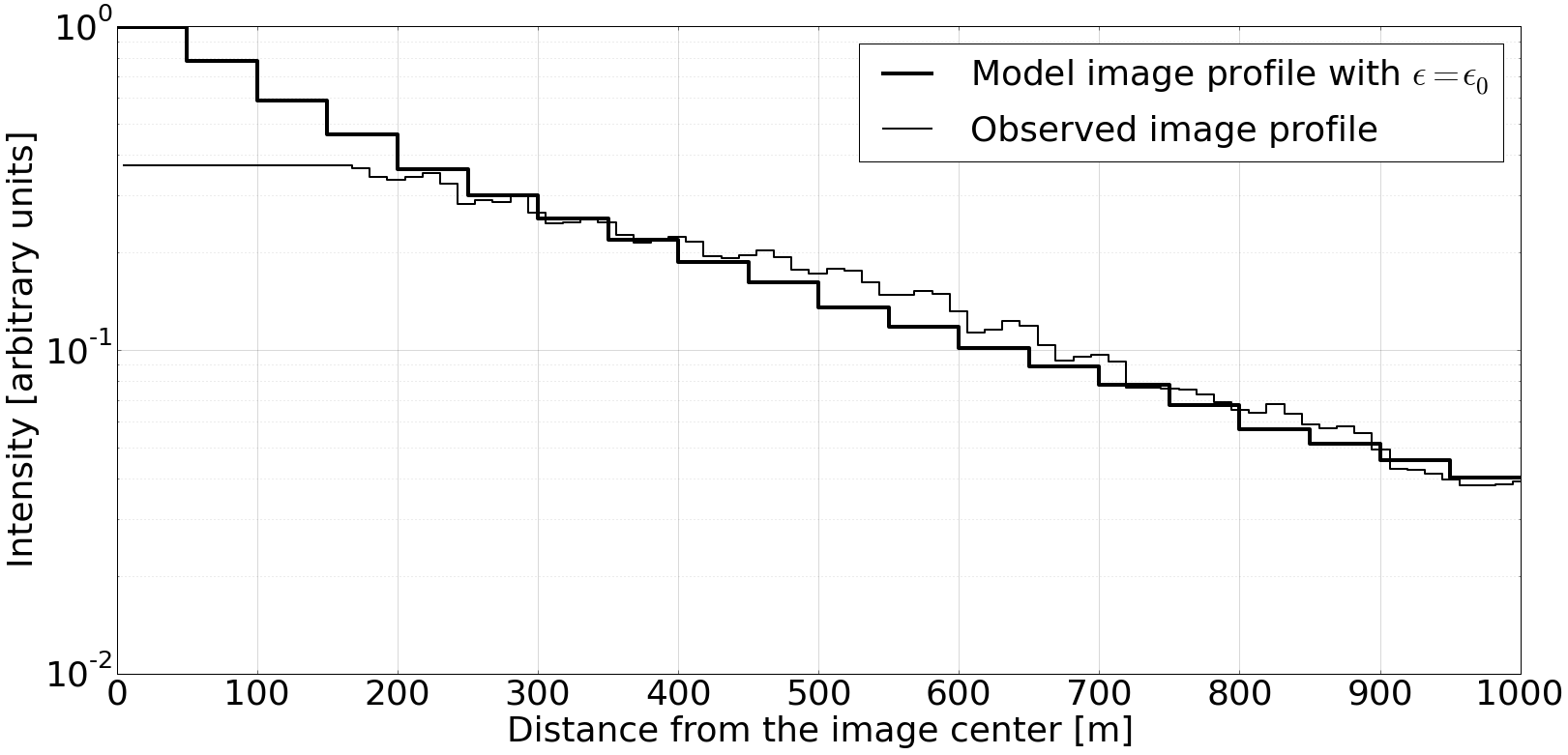}
  \caption{The simulated halo image and the intensity profile using a model where the scattering parameter is a constant $\epsilon=\epsilon_0$ (see equation \ref{eq:scattering}). An improvement is visible in matching the modeled and measured brightness profiles compared to the canonical model in Fig. \ref{fig:canonical_intensity}. }
  \label{fig:epsilon}
\end{figure}

\begin{figure*}
  \centering
    \includegraphics[width=0.4\textwidth]{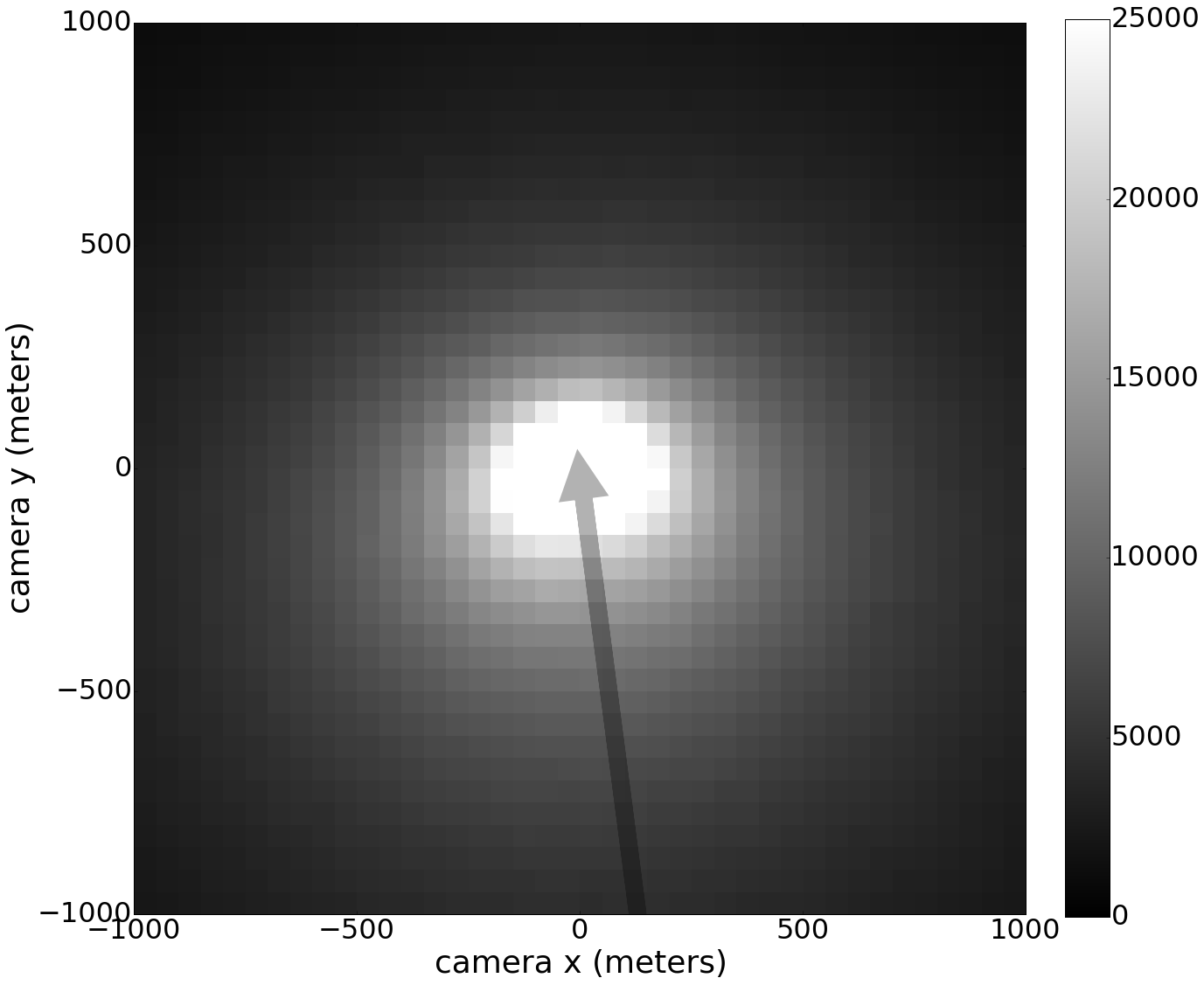}
    \includegraphics[width=0.4\textwidth]{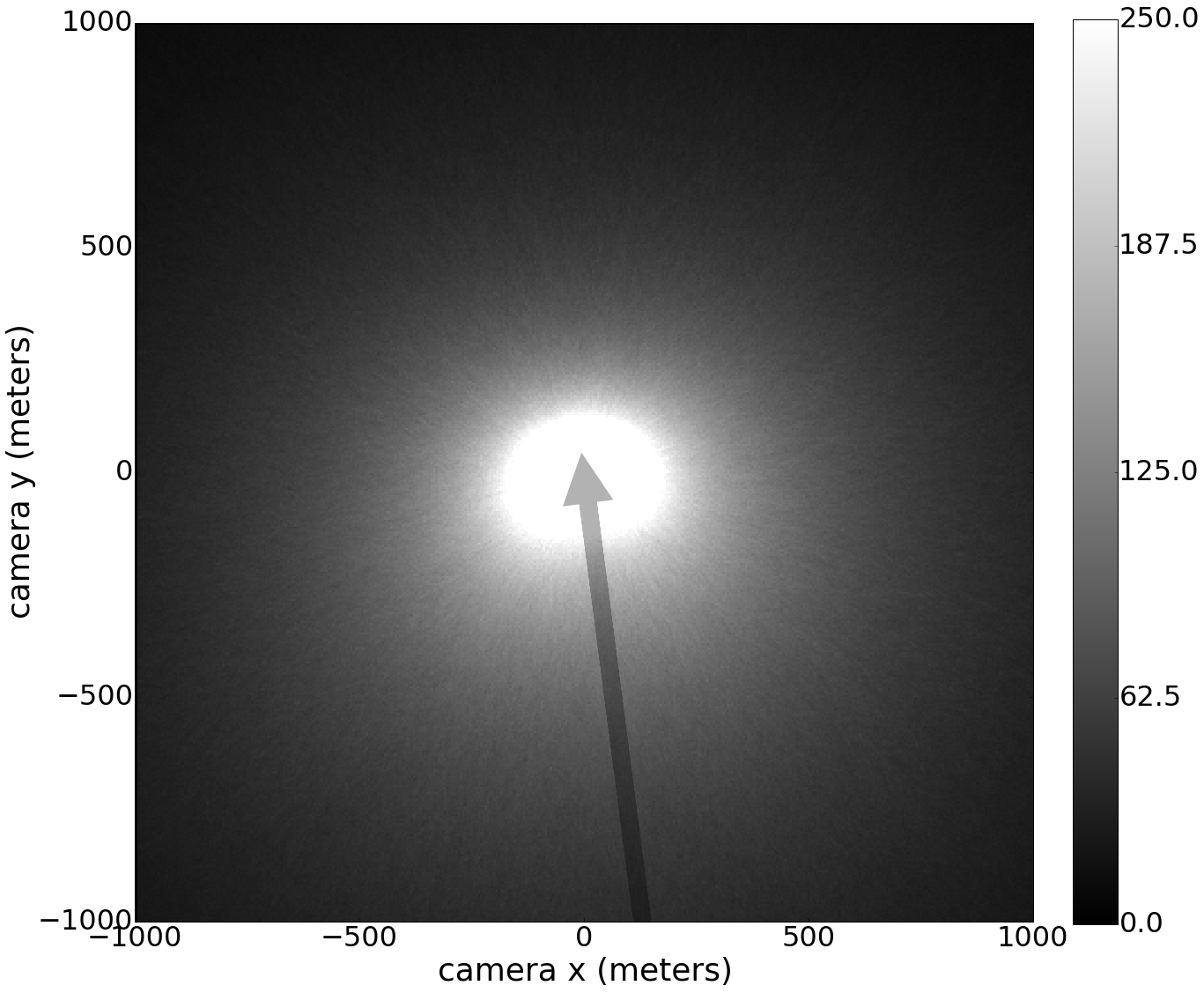}
    \includegraphics[width=0.4\textwidth]{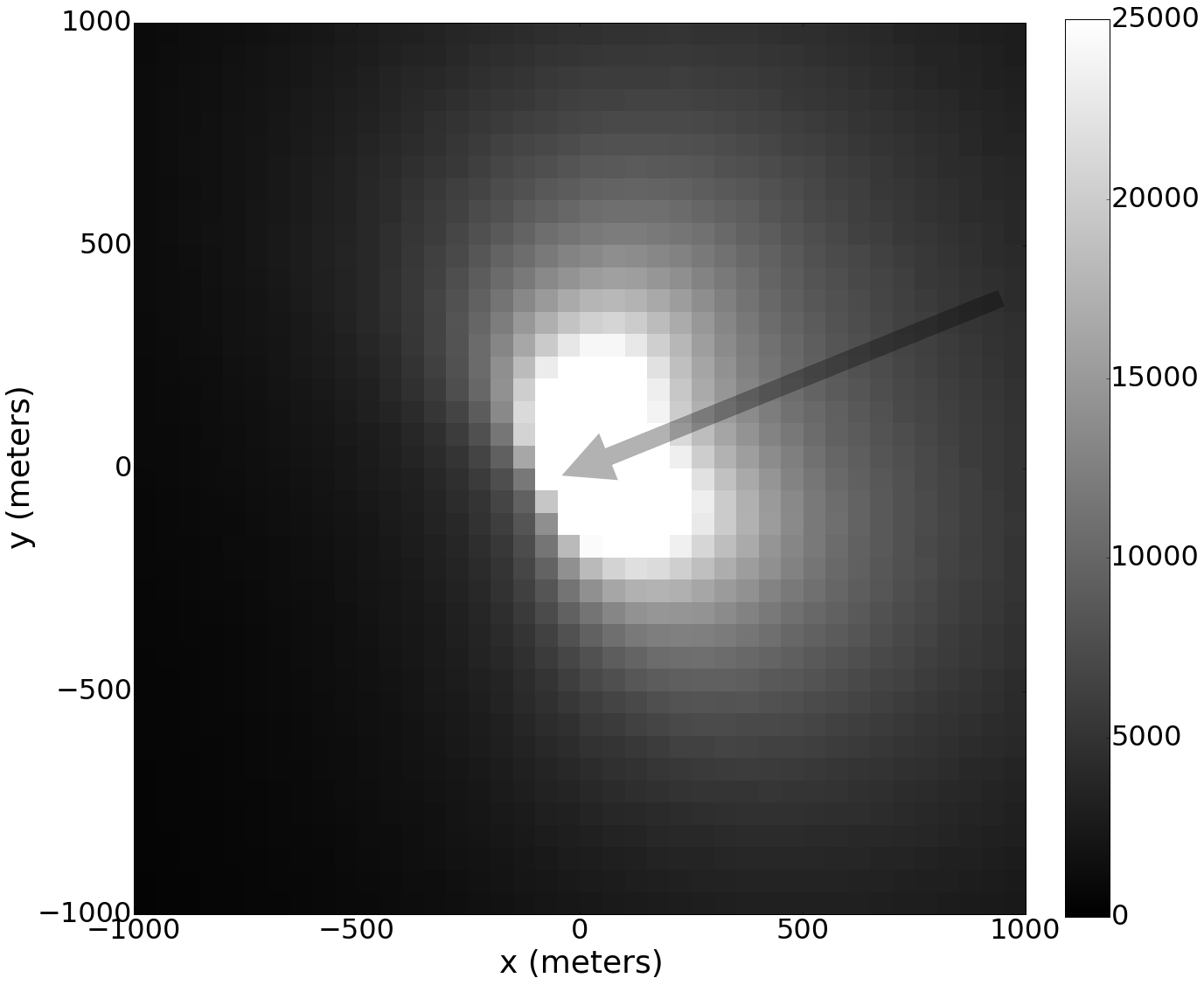}
    \includegraphics[width=0.4\textwidth]{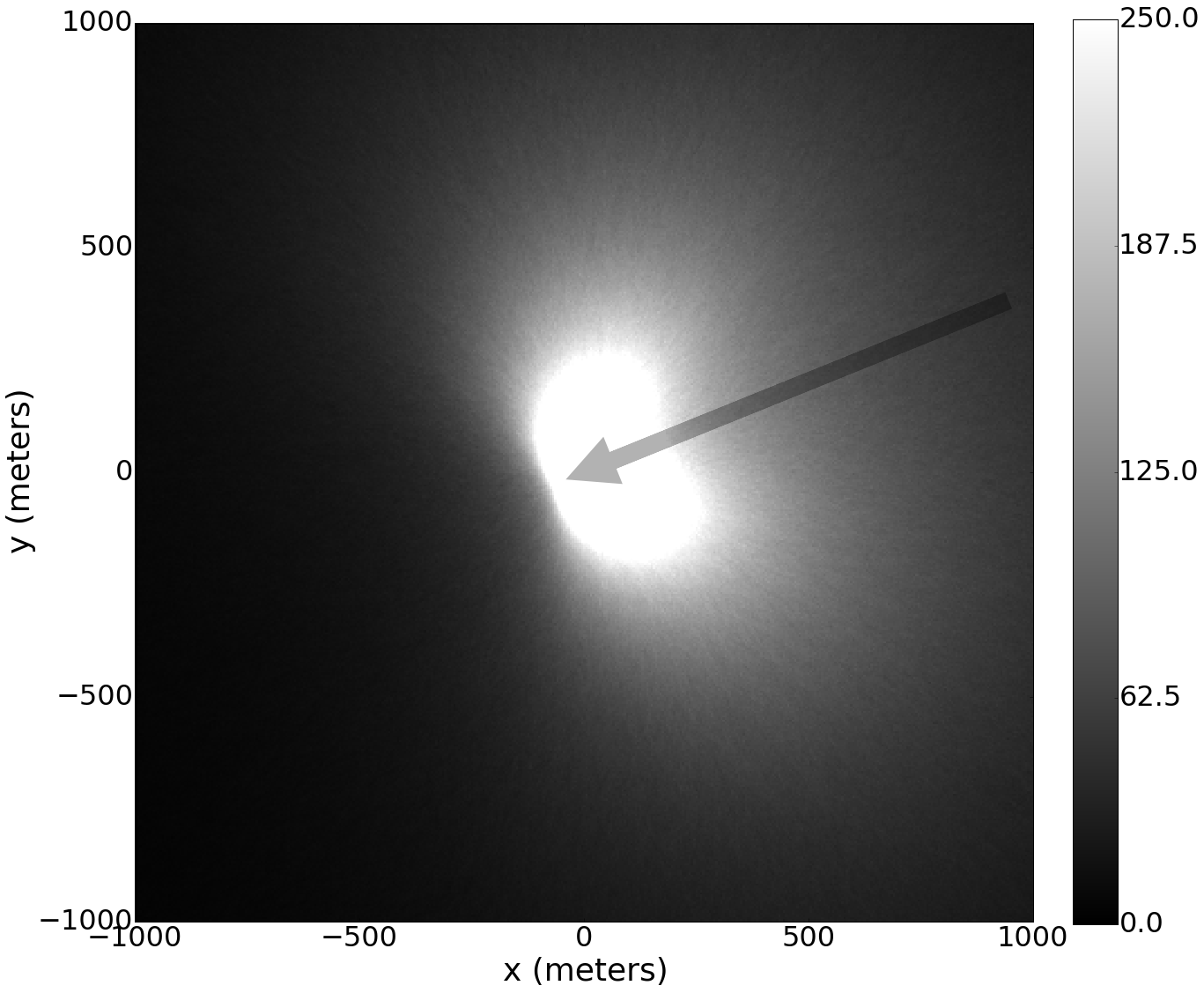}
  \caption{Simulated meteor halo structure at 105 km altitude that was detected around a Leonid meteor \citep{REFFStenbaekNielsen}. The brightness scale is in arbitrary units. Simulation used $10^6$ particles ejected from the meteor head. Brightness from the meteor itself is not shown, but gray arrows mark the meteor's position and direction. Images in the upper row show the halo from the view point of the camera that detected the halo, while the lower row shows the halo from a sideview direction perpendicular to the meteor flight trajectory. The left column images have the pixel size similar to the camera and the right column images have a high resolution that shows the halo's fine structure. The emission is a product of proton/hydrogen collisions with the neutral atmosphere, after the protons become ejected from the meteor head due to the meteor's electric field.}
  \label{Fig:mainFig}
\end{figure*}

\begin{figure}
  \centering
  \includegraphics[width=0.49\textwidth]{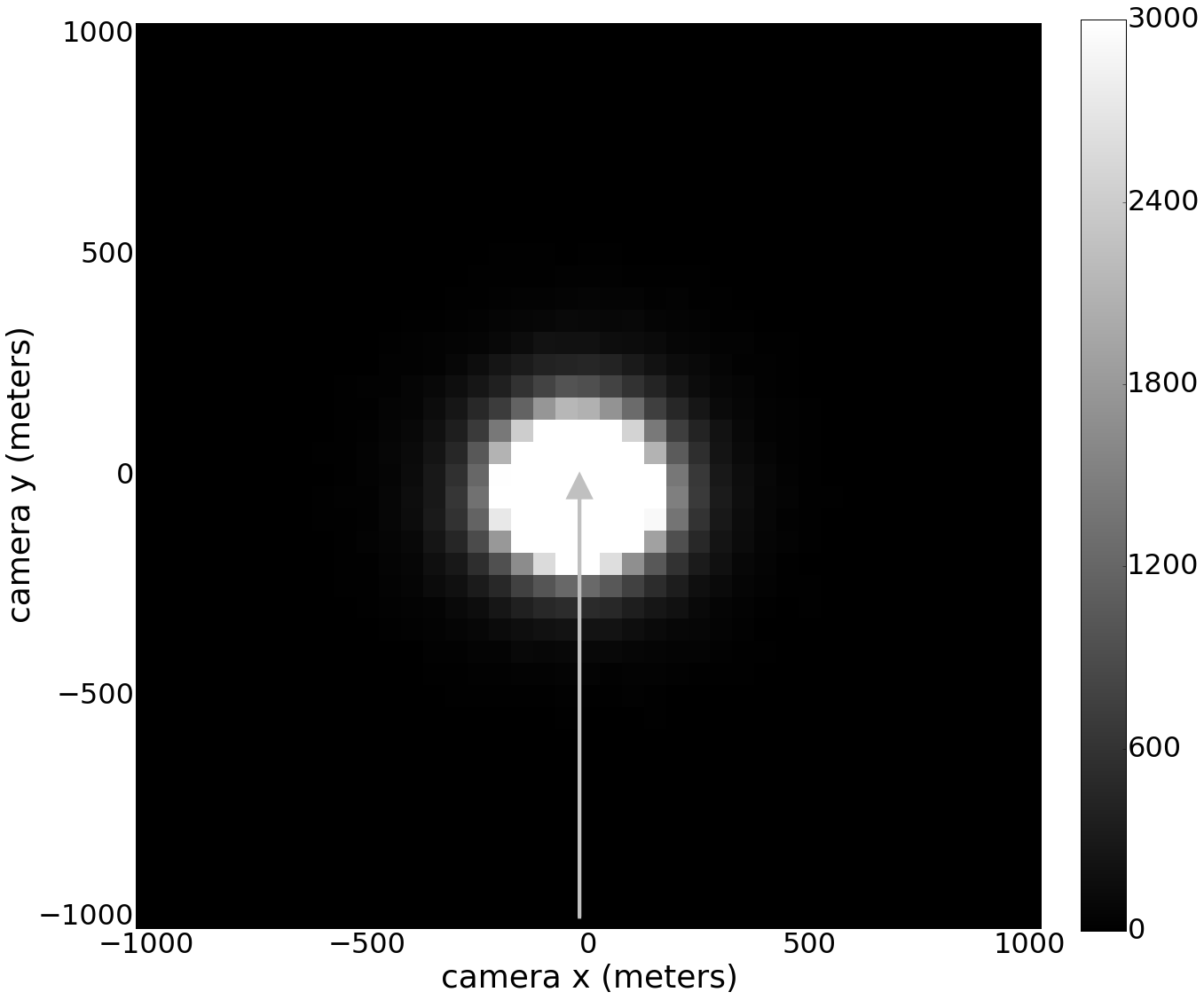}
  \caption{A simulated halo image using the model of UV photon absorption under the (unrealistic) assumption of immediate emission of visible photons (UV photons should actually be emitted on timescales slower than the frame rate of the high-speed camera). The brightness level is adjusted to give saturated pixels up to about 200 m from the meteor center to match the observed image. The model intensity drops so fast with distance that there is no detectable halo brightness at larger distances from the meteor. Hence, this excludes UV excitation as the source of halo brightness. The model used 1 million UV photons emitted along directions from 0$^\circ$ to 120$^\circ$ relative to the meteor direction. }
  \label{fig:UVimage}
\end{figure}

For a particle traveling with the speed $\text{v}_p$, the location of its collision is derived from the time of flight $T_{coll}$ through the atmosphere
\begin{equation}\label{Path-integral}
 -\ln \Re = \int\limits_0^{T_{coll}} \text{v}_p(t) \sum\limits_i Q_i(\text{v}_p) n_{atm}(i)\,\,  dt
\end{equation}
where $\Re\in (0,1]$ is a random number and $i\in\{O,O_2,N_2\}$. The collisional cross section $Q_i$ is the total sum of all cross sections processes for either a proton or a hydrogen atom projectile and it depends on the projectile's kinetic energy, i.e. on $\text{v}_p$. This equation can be solved only by numerical integration, since the particle speed $\text{v}_p$ is changing in time due to the influence of electric field. For this purpose we use a simple trapezium rule with the integration time step $\Delta t=10^{-7}$ s. The collision position is shifted by $-kT_{exp}\overrightarrow{\text{v}}_m$, where $k$ is the number of exposure times $T_{exp}$ that passed since the start of the exposure. In other words, when the particle flight (including its ejection time $\Re\,T_{exp}$) exceeds the exposure time of the camera frame, it can be considered as a particle from an older ejection that extended its collisional lifetime into this exposure of our interest.

When a collision happens the collision target is selected randomly form the relative abundances of N$_2$, O$_2$ and O. After that the type of collision (ionization, excitations, charge exchange or electron stripping) is also selected randomly from the relative values of their cross sections. Irrespective of the collision type, we memorize the collision location and assign it the relative value of the nitrogen excitation cross section $Q_{exc}(N_2)/Q_{N_2}$ to speed up the calculation. This is done because our particles are actually statistical representatives of a bunch of particles, where this fraction of them might have participated in the nitrogen excitation. These values are added to the pixels of virtual cameras positioned around the meteor.

A proton quickly reaches 1 keV after ejection and then its path is traced until its energy drops below 1keV or its distance from the meteor exceeds 10 km. Even though some projectiles can still induce nitrogen excitation at these distances, they are spread within such a big volume that their contribution to the image surface brightness becomes negligible. Fig. \ref{fig:energy} shows an example of the proton/hydrogen energy at various distances from the meteor. The maximum energy reached in our simulation is about 4keV. The right panel in the figure shows the excitation cross section relevant for photon production. It illustrates the relative importance of different projectile energies in light production.

A sketch of our model is shown in Fig. \ref{fig:sketch}, while the pseudocode that describes the overall simulation procedure is as follows:

{\small
  \begin{algorithmic}[1]
  \STATE Set initial parameters
    \FOR {loop over ejected particles}
      \STATE Set the initial particle position and velocity
      \REPEAT
        \STATE Set the collision integral value $I= -\ln \emph{R}$
        \REPEAT
          \STATE Integrate the collision integral in equation \ref{Path-integral}
          \STATE Find the electric field
          \STATE Update particle position and velocity
        \UNTIL {the integral becomes equal to $I$}
        \STATE Find $k$ and shift the collision location
        \STATE Update camera pixel values for this location
        \STATE Find the collision type and atmospheric species
        \STATE Switch between proton and hydrogen if required by the collision type
        \STATE Find the energy reduction for this collision type
        \STATE Find new path direction for the particle
        \STATE Update the particle velocity from the new direction and reduced kinetic energy
      \UNTIL {the particle exceeds energy or position limits }
    \ENDFOR
    \medskip
  \end{algorithmic}
}

\subsection{Simulation results}
\label{subsec:results}

The detected halo presented by \cite{REFFStenbaekNielsen} in their Figure~1 extends hundreds of meters from the meteor head, with a slight brightness even at 1 km distance. On this scale our model manages to keep this level of brightness, as seen in Fig. \ref{fig:canonical_intensity}. The inner 200 meters around the meteor lack comparison because of pixel saturation in the observed images. There is also observation by  \cite{REFFStenbaekNielsen} in their Figure~6 supposedly showing a structure smaller than 200 meters, but that figure lacks any supporting technical description about the brightness reproduction and pixel size. If we use in that case the image brightness as prima facie grayscale values in the same way as in their Figure~1, and accept the image scale shown in their Figure~6 to be correct, then our model can reproduce the halo brightness profile in their Figure~6 if $p=0.4$. However, due to uncertainties with the observed brightness reproduction we do not investigate this in any further detail.

Since our canonical model is based on $E_0=$300 V/m and $p=$2/3, we explore how changes in $E_0$ and $p$ influence the proton/hydrogen transport and, consequently, the halo image. If we neglect collisions for a moment then we can derive the analytic solution for the particle energy (in eV) from the particle acceleration $d\text{v}/dt=(d\text{v}/dr)(dr/dt)=0.5d\text{v}^2/dt$ being equal to the electric force $(|e|E_0/m_p)/(r/r_0)^p$, where $m_p$ is the proton mass. In such a case the solution for the proton/hydrogen energy at distance $r$ from the meteoroid is $E_0(r^{1-p}-r_0^{1-p})/(1-p)$ when $p\neq1$ and $E_0 \ln (r/r_0)$ when $p=1$. We see that the energy depends linearly on $E_0$, while changes in $p$ from 2/3 to 1.5 yield energy drop by a factor of $\sim$10. This gives us a general sense of trends. 

In Fig. \ref{fig:change_p_E0} we show how our simulated particle energy changes with distance when collisions are included. The main trends with $E_0$ and $p$ remain, but now collisions slow down particles as they move away from the meteoroid. The figure shows that $p=2/3$ keeps the particle energy almost constant within 1 km distance from the meteoroid, while larger values of $p$ lead to a declining energy with distance. The right panel in Fig. \ref{fig:change_p_E0} shows the nitrogen excitation cross sections, which is equivalent to the relative probability of photon emission. This analysis implies two major consequences on the halo image appearance: reduced $E_0$ by a factor of 2 leads to lower halo brightness by a factor of $\sim$10, while increased $p$ creates a smaller halo image since particles loose their collision energy with distance faster than in the canonical model. This is then modulated with the sensitivity and resolution of the imaging detector.

Since our model is not an ideal fit to the observations, we tried to deviate from the canonical model. In Fig. \ref{fig:p=075} we show how a slight change to the power law from p=2/3 to p=3/4 improves the fit. And if we allow for experimentation with the proton scattering parameter $\epsilon$ then we have a situation like in Fig. \ref{fig:epsilon} where the fit is improved by keeping $\epsilon$ constant. Obviously, there is some degeneracy in the modeling parameters because of the way how we configured our approximate electric field.

\subsection{The optimal modeling result}
\label{sec:optimal}

The result of our halo simulation is shown in Fig. \ref{Fig:mainFig} from two viewpoints - the camera that detected the halo around a Leonid meteor and the sideview that provides a better insight into the halo structure. The meteor itself is not shown, but the arrow marks the position of the meteor head and meteor trail. The image reconstructs the size of the detected halo. The brightness profile of the halo perpendicular to the meteoroid velocity vector is shown in Fig. \ref{fig:canonical_intensity} and compared with the observed brightness profile. 

Our attempts to reconstruct the halo image by ejecting protons into various solid angles positioned along the direction of meteor flight resulted in an interesting conclusion that ejection angles should prefer backward directions (Appendix \ref{sec:angles}). The reason for this angle exclusion mechanism is the small viewing angle of about 30$^\circ$ between the camera direction and the meteor line of flight, which makes axially-symmetric halo structures appear very symmetric and positioned too far ahead of the meteoroid. The solution was to eject protons mainly into backward directions. The search for the best fit to the brightness profile yielded a choice where protons are ejected into directions between $\vartheta_0=60^\circ$ and $\vartheta_1=150^\circ$ relative to the direction of meteor flight, with a probability shown in equation \ref{eq:ejectangle}. These angles should be considered only as a rough approximation since we are dealing with an approximate model of the meteor head plasma configuration and ionospheric electric field geometry and screening.

It is interesting to note that a numerical simulation of radiative gasdynamic model of physically and chemically nonequilibrium flow around an ablating meteoroid yields an enhanced hydrogen production predominantly in the backward sidelobes relative to the meteoroid \cite{REFFSurzhikov}. This is in agreement with our constraint on the ejection angles and with the expectation that the flow momentum will carry plasma atoms out of the back and sides of the meteor head.

\subsection{UV model}
\label{sec:UVmodel}

We also show a model of UV propagation in Fig. \ref{fig:canonical_intensity} using a simplified favorable assumption that the visible photons from recombination are emitted immediately at the point of absorption of UV photons. Recombination is actually a slow process in the ionospheric plasma and it would create a delay in photon emission, which turns the halo image into a droplet shape, unlike the detected shock-like structure, but for the sake of the argument we ignore this effect. We use absorption cross sections in the 700-750${\mathring{\text{A}}}$ wavelength range \citep{REFFSchunk}, as a typical sample of UV cross sections. The numerical procedure is similar to the proton ejection model, except that photons are absorbed and removed from the simulation at the first interaction with the neutral atmospheric species. 

We see in Fig. \ref{fig:canonical_intensity} that such a UV model cannot explain the halo even with this highly favorable (albeit unrealistic) assumption of immediate emission. The UV gets absorbed too quickly in the atmosphere and the brightness falls sharply with the distance, as shown in Fig. \ref{fig:UVimage}. With this UV simulation we also tested the suggestion that UV photons escape into the atmosphere only within a limited range of azimuthal angles (i.e. backward directions are blocked by the meteor) \citep{REFFJenniskensD}, but Fig. \ref{fig:UVimage} shows that this results in a highly symmetric image (because of a small angle between the meteor direction and the line of sight), unlike the observed shock like structure. Notice, however, that we do not exclude the existence of precursor UV photons that would ionize nearby neutral atmosphere, but we show how they do not play a role in the production of a large scale halo around the meteor.

\section{Conclusions}

The existence of a kilometre-scale halo caused by a corona discharge from a charged meteor had been speculated already in the 60's \citep{REFFDokuchayev}, but weak theoretical arguments and the lack of convicting observational confirmation of charged meteors had kept this phenomenon on the fringe of meteor astronomy. Our model of charged meteor plasma infers that detected meteor halo \citep{REFFStenbaekNielsen,Jenniskens2002}, oscillations of meteor head plasma \citep{REFFKero}, and fragmentation of meteoroids that result in large transverse speeds \citep{REFFStokan} could be all manifestations of charged meteors. Moreover, the spatial size of produced ionospheric irregularity of enhanced ionization is such that it is maybe a trigger of various other phenomena. For example, the first instrumental detection of electrophonic sounds performed during the 1998 Leonid meteor storm \citep{REFFZgrablic} pointed to the existence of a larger ionospheric disturbance triggered by meteors. Also, a kilometer-scale plasma irregularities at about 90 km altitude are identified recently as precursors of sprite streamer initiation, with meteors as a possible origin of these irregularities \citep{REFFQin}.

Phenomena like fast jet-like features from meteors \citep{REFFSpurny,REFFLeBlanc} might also be related to strong local electric fields. In jets from high-altitude meteors \citep[above $\sim$130 km;][]{REFFSpurny2000} the magnetic and electric fields must play an important role as the meteor plasma is of a low density where both the electrons and ions are magnetized. Also, the role of possible trail currents \citep{REFFDimant} in enhancing the meteor head electric field is not explored in detail. This and other processes could boost the meteor head charging. For example, the magnetic Reynolds number indicates that below 100 km altitude the magnetic field might become compressed in front of the meteoroid (stagnation zone) and enhance charge separation processes. In general, we need to explore meteor plasma microphysics in much more detail, with external magnetic and electric fields included, while concurrently put more effort into observations of various poorly understood meteor related phenomena.

\section*{Acknowledgements}

The work on computational simulations was performed at the Physics Department, University of Split, and included their CPU/GPU cluster Hybrid, financed by the National Foundation for Science, Higher Education and Technological Development of the Republic of Croatia. The authors also acknowledge the Technology Innovation Centre Me\dj{}imurje for additional computational resources. D.V. acknowledges membership in the network supported by the COST Action TD1403 Big Data Era in Sky and Earth Observation.





\appendix
\renewcommand\thefigure{A\arabic{figure}}   

\section{Magnetic Reynolds number in meteor plasma}
\label{sec:magnetic}

Magnetic Reynolds number $R_m$ is defined as $R_m = \mu_0 \sigma_0^m L_0 \text{v}_0$, where $\mu_0$ is the magnetic permeability, $\sigma_0^m$ is the plasma conductivity, $L_0$ is the characteristic length of the meteor and $v_0$ is the characteristic velocity. The meteor head size in our case is $L_0\sim 1$m and $\text{v}_0=\text{v}_m$ is the meteor velocity. The conductivity is dominated by elections, which display the highest conductivity parallel to the magnetic field  $\sigma_0^m=n_e^m e^2 / m_e \nu_e$, where $m_e$ is the electron mass and $\nu_e$ is the collision frequency. Measurements indicate the upper meteor electron density of $n_e^m\sim 10^{16}$ m$^{-3}$ for meteors at 105 km altitude \citep{REFFClose}.

The collision frequency can be estimated from its basic definition as $\nu_e=n_m Q_e \langle v \rangle$, where $n_m$ is the number density of neutrals in the meteor plasma, $Q_e$ is the collisional cross section and $\langle v \rangle= 6\times 10^3 \sqrt{T_e}$ is the average thermal speed of electrons (in units of m/s) of temperature $T_e$ (in Kelvins). The measured equilibrium temperatures in meteor plasma remain in a narrow range around $\sim$4,500 K for a wide range of entry speed and initial mass \citep{REFFJenniskensB}. The electrons in meteor plasma collide mostly with atomic species, with $Q_e\sim 5{\mathring{\text{A}}}^2$ as an approximate average collisional cross section \citep[at $\sim$1 eV; e.g. see ][]{REFItikawa}. 

The plasma density can be estimated by comparing dynamics pressure of incoming air and the thermal pressure of compressed meteoroid plasma $\eta m_{atm}n_{atm}\text{v}_m^2/2=n_m k T_m$, where $m_{atm}$ is the average atmospheric molecular mass and $n_{atm}$ is the atmosphere number density. The plasma temperature is assumed in equilibrium as $T_m=T_e$. The dimensionless constant $\eta<1$ is introduced to accommodate for a fast drop in pressure outside the stagnation zone, as seen in numerical simulations \citep{REFFBoyd,REFFSurzhikov}.

Conditions at 105 km altitude are $m_{atm}=4.6\times 10^{-26}$ kg and $n_{atm}=4.4\times 10^{18}$ m$^{-3}$, which gives $n_m=\eta \times 10^{22}$ m$^{-3}$. This agrees well with the range of densities in numerically simulated meteors \citep{REFFBoyd}, which show $\eta < 10^{-2}$ in the relaxation zone on the sides and behind the meteoroid \citep{REFFBoyd,REFFSurzhikov}.  Using these values we derive the collision frequency $\nu_e=2\eta \times 10^{8}$ Hz. Notice that this makes $\nu_e < \omega_{ce}$.

Plugging all these values into the equation for $R_m$ gives $R_m\leq 0.25 \eta < 1$. This shows that at these altitudes the magnetic field lines are not frozen into the meteor plasma. However, at altitudes below 100 km the maximum meteor head plasma density increases to $n_e^m\sim 10^{18}$ m$^{-3}$, which would enable dragging of the magnetic field lines by the meteor plasma in the meteoroid vicinity (where the plasma has its maximum density).

\begin{figure*}
  \centering
  \includegraphics[width=0.4\textwidth]{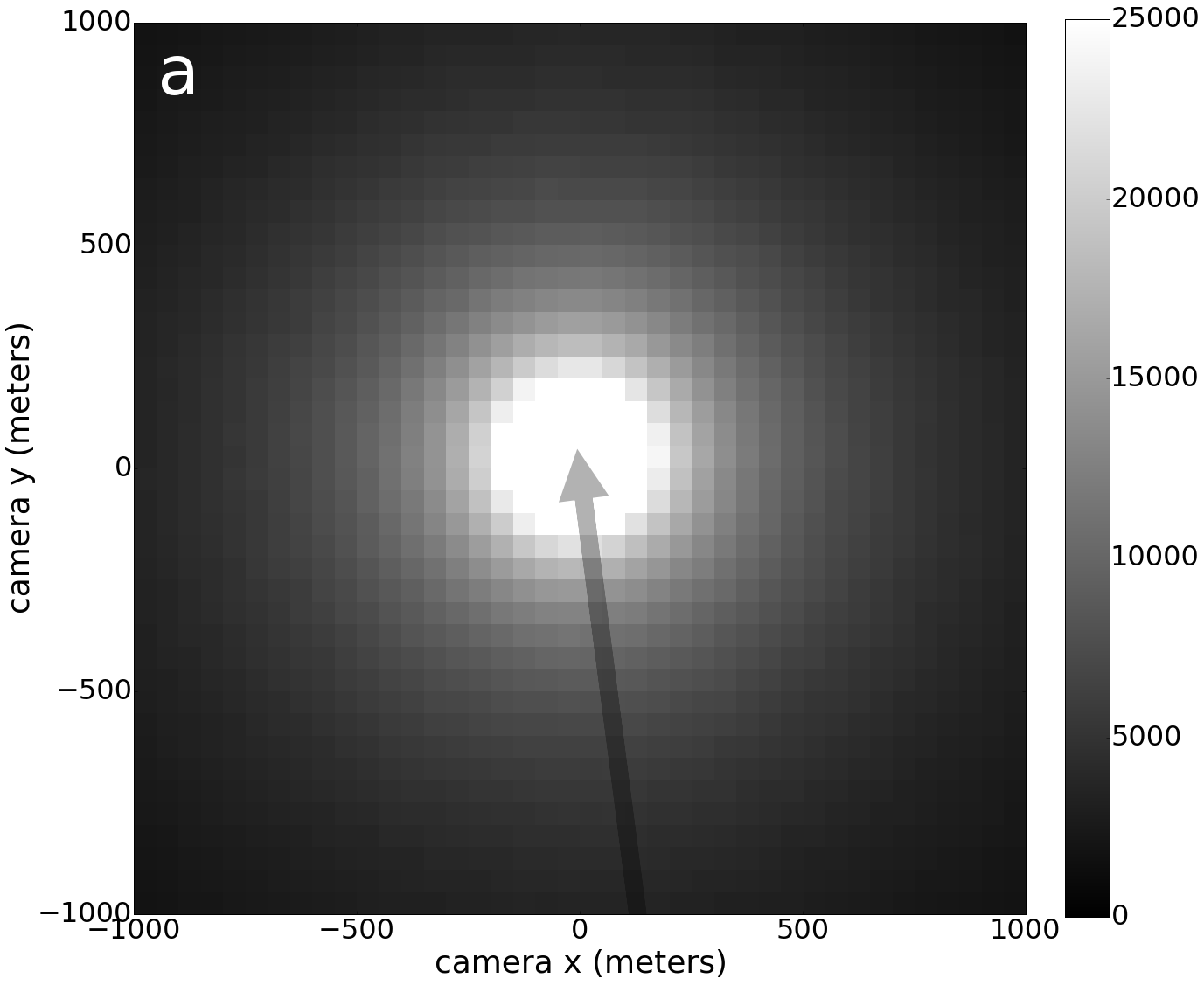}
  \includegraphics[width=0.4\textwidth]{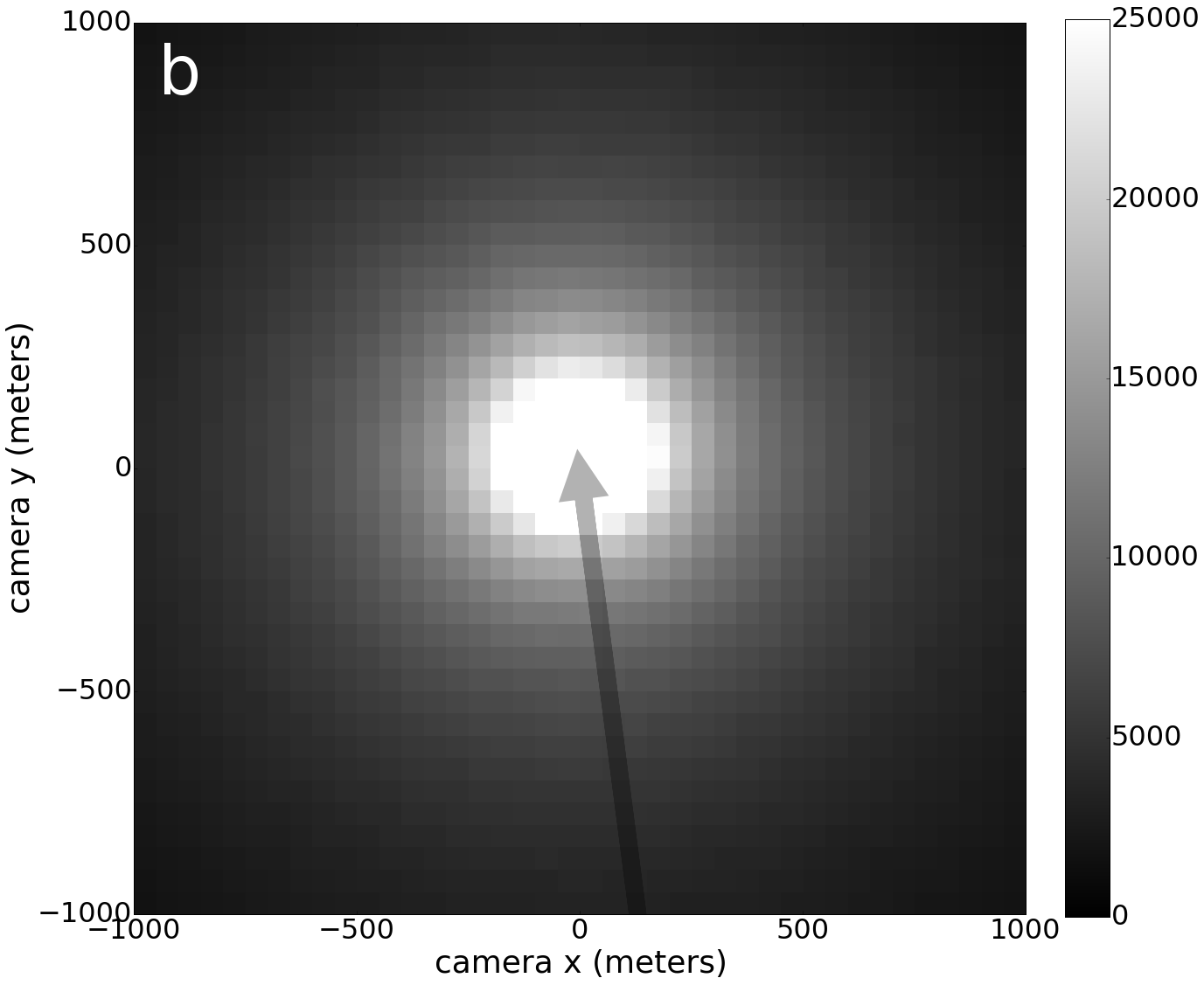}
  \includegraphics[width=0.4\textwidth]{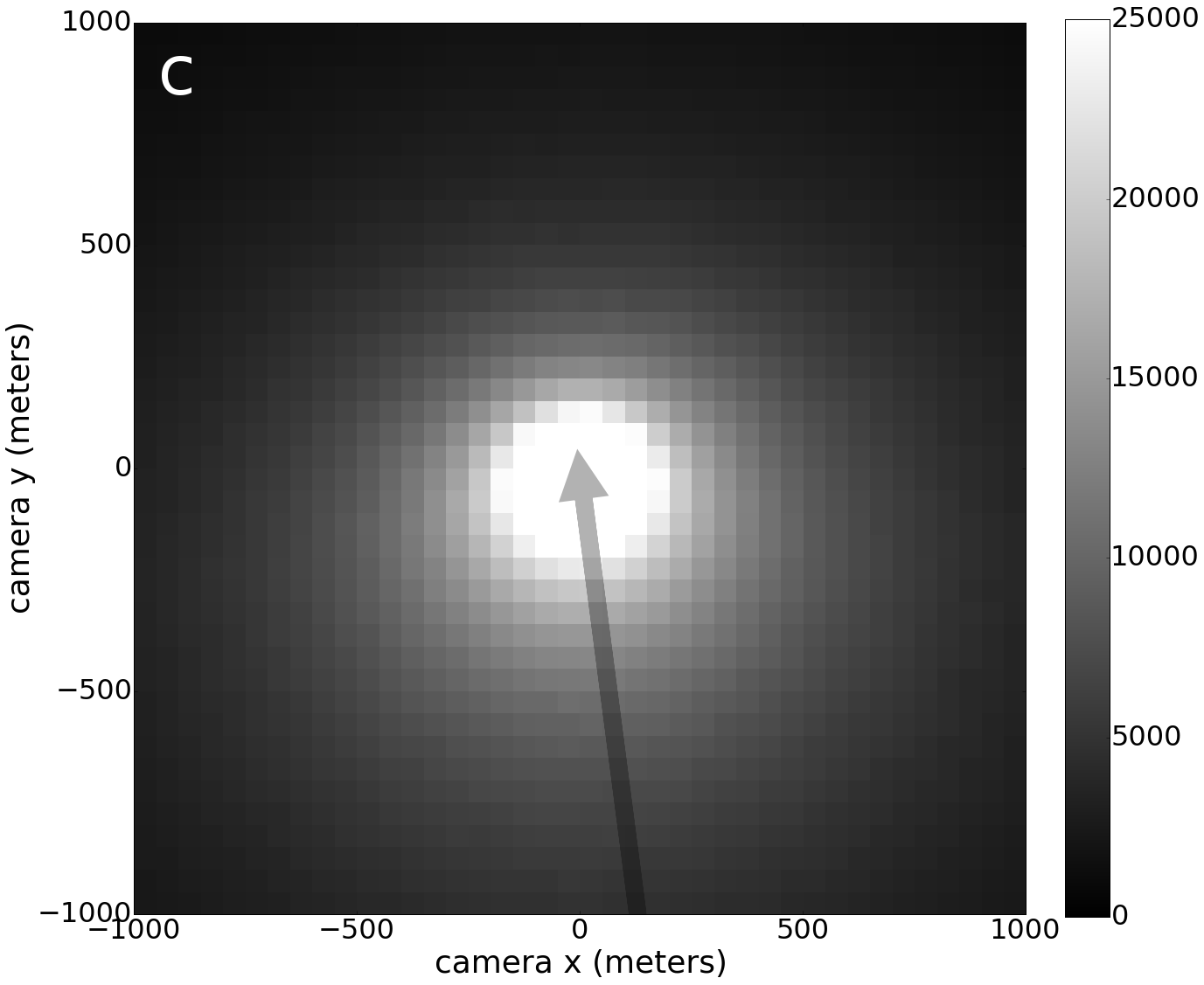}
  \includegraphics[width=0.4\textwidth]{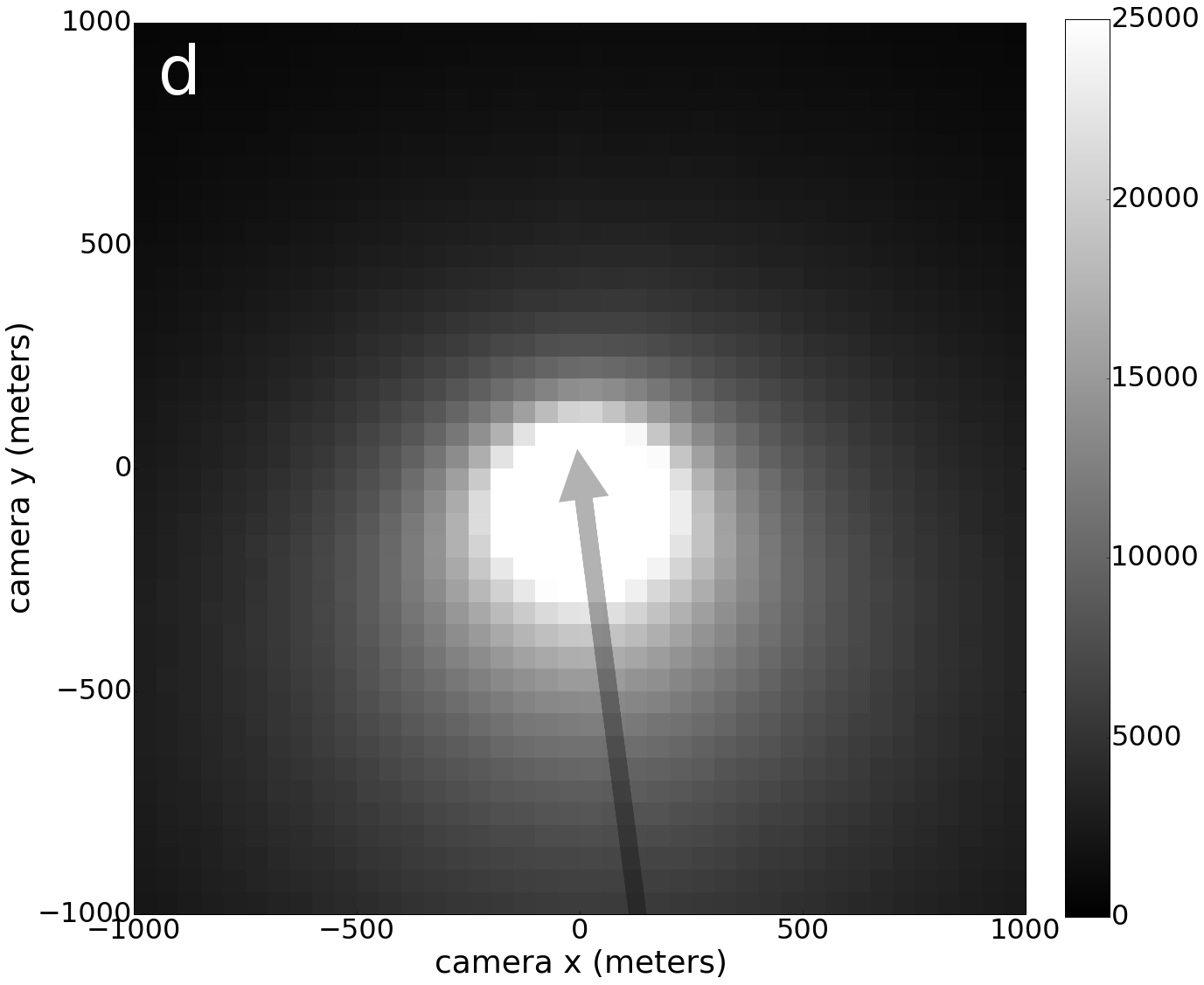}
  \caption{Simulated images of the canonical Leonid meteor model calculated with proton ejection angles between ($\vartheta_0$,$\vartheta_1$) distributed according to the equation \ref{eq:ejectangle}. Panels {\it a},{\it b},{\it c} and {\it d} have (0$^\circ$,180$^\circ$), (0$^\circ$,160$^\circ$), (90$^\circ$,180$^\circ$) and (120$^\circ$,180$^\circ$), respectively. It is evident that matching model images to the observed halo requires an emphasis on backward ejection angles.}
  \label{fig:angles}
\end{figure*}

\section{Hall current and Cowling conductivity}
\label{Cowling}

In this approximative analysis we use $\overrightarrow{\text{v}}_m \perp \overrightarrow{B}$, which is the most interesting configuration in our discussion on meteor halo and it also produces the strongest Hall currents. The Hall current stops when the electrostatic secondary polarization field $E_H=E_m \sigma^m_H/\sigma^m_P$ is generated perpendicular to $\overrightarrow{\text{v}}_m$ and $\overrightarrow{B}$. We are interested in situations when the meteor is surrounded by a diffuse shock that has $\nu_e < \omega_{ce}$, which gives the meteor Hall conductivity of $\sigma^m_H\approx \sigma^m_0 \nu_e / \omega_{ce}$ and the meteor Pedersen conductivity of $\sigma^m_P\approx \sigma^m_0 \nu^2_e / \omega^2_{ce}$. Under such conditions the conductivity along the direction of meteor path, which is responsible for the neutralization current, would be enhanced to the Cowling conductivity $\sigma^m_C= \sigma^m_P(1+(\sigma^m_H/\sigma^m_P)^2)\approx \sigma^m_0$. But such a high conductivity would drastically reduce $E_m$, which in turn would reduce the Hall current and there would be no time to induce the Cowling conductivity. Hence, there must be some intermediate equilibrium situation that balances those two trends.  

This equilibrium can be derived by considering timescales needed for moving charge through the meteor plasma. The Cowling effect is relevant if there is enough time for the Hall current to establish $E_H$. The timescale of transversely moving charge is $\tau_H= L_0/\,\overline{\text{v}}_H$, where $L_0$ is the meteor width perpendicular to its direction of flight and $\overline{\text{v}}_H$ is the mean speed of charge within the Hall current. The mean speed can be derived from the definition of polarization current $e n^m_e\,\overline{\text{v}}_H = \sigma^m_H E_m$. We derive $\overline{\text{v}}_H=(e/m_e)E_m\omega_{ce}/(\nu^2_e + \omega^2_{ce})\sim e E_m/m_e\omega_{ce}$ and combine it with equation \ref{eq:Em} to obtain $\tau_H\sim L_0\chi\omega_{ce}/\text{v}_m\nu_e$. This transverse motion takes place at distance $L$ from the tip of the meteor head and it generates the charge separation within the meteor flight timescale $\tau_m=L/\text{v}_m$. When we compare those two timescales such that $\tau_H\approx\tau_m$, we obtain the equilibrium value of conductivity along the meteor path $\chi\sim (\nu_e/\omega_{ce})(L/L_0)$.

\section{Electric field screening}
\label{sec:screening}

The ionospheric plasma reacts to the positive electric field from a meteor by readjusting the distribution of its charged particles to neutralize the field. We can view this as an accumulation of negative charge closer to the positively charged meteor. We work with a simple point source electric field geometry and homogenous plasma currents. A more realistic electric field is probably not like a point source and the ionospheric currents are influenced by the anisotropic electron mobility. Thus, our description of the screening is an approximation to be used a starting point for exploring the meteor halo structure.

We assume a steady state process of screening, which yields the following steps:

1. A steady state flow of negative charge of density $n_-(r)$ gives $\partial n_-(r)/\partial t=0$.

2. The charge conservation in spherical geometry is
\begin{equation}
\frac{1}{r^2}\frac{\partial}{\partial r}\left( r^2 n_-(r) \text{v}_-(r) \right)=0
\end{equation}
where $\text{v}_-(r)$ is the ionospheric (average) radial charge velocity during the screening process. This gives $\text{v}_-(r)\propto 1/(r^2 n_-(r))$.

3. The radial component of the momentum equation for ionospheric charge dynamics around a meteor gives $eE'(r)=m_e\nu_{en}\text{v}_-(r)$, where $E'(r)$ is the electric field around the meteor and $\nu_{en}$ is the electron-neutral collision frequency. The collision frequency can be approximated with $\nu_{en}=n_n\sigma_{en}\text{v}_-(r)$, where $n_n$ is the density of neutral particles and $\sigma_{en}$ is the collisional cross section. If we plug this into the momentum equation, and combine it with the equation in step 2 equation for $\text{v}_-(r)$, we get
\begin{equation}
E'(r) \propto \frac{1}{r^4n_-^2(r)}
\end{equation}

4. The radial component of the Gauss law gives
\begin{equation}
 \frac{1}{r^2}  \frac{\partial}{\partial r}\left( r^2  E'(r)\right) \propto n_-(r)
\end{equation}

5. Now we have to find the electric field that satisfies both steps 3 and 4. Let us assume $E'(r)\propto r^{-p}$. From the step 4 we get $n_-(r)\propto r^{-1-p}$ and from the step 3 we get $n_-(r)\propto r^{p/2 - 2}$. These can be satisfied if $p/2-2=-1-p$, which gives $p=2/3$.

\section{Proton ejection angles}
\label{sec:angles}

Changes in the range of proton ejection angles defined by equation \ref{eq:ejectangle} impact the halo shape. With this equation we defined an equal probability of selecting polar angles between the limits $\vartheta_0$ and $\vartheta_1$. In Fig. \ref{fig:angles} we show examples of simulated halo images produced with different combinations of ($\vartheta_0$,$\vartheta_1$) angles.


\bsp	
\label{lastpage}

\begin{thebibliography}{99}

\bibitem[\protect\citeauthoryear{Basu et al.}{1987}]{REFFBasu87}
 Basu B., et al.\ 1987, J. Geophys. Res., 92, 5920

\bibitem[\protect\citeauthoryear{Basu et al.}{1993}]{REFFBasu93}
 Basu B., et al.\ 1993, J. Geophys. Res., 98, 21517

\bibitem[\protect\citeauthoryear{Beech \& Foschini}{1999}]{REFFBeech}
 Beech M., Foschini L.\ 1999, Astron. Astrophys., 345, L27

\bibitem[\protect\citeauthoryear{Borovi\v{c}ka}{1994}]{REFBorovicka94}
 Borovi\v{c}ka J.\ 1994, Planet. Space Sci., 42, 145

\bibitem[\protect\citeauthoryear{Borovi\v{c}ka et al.}{1999}]{REFBorovicka99}
 Borovi\v{c}ka J., Stork R., Bocek J.\ 1999, Meteorit. Planet. Sci., 34, 987

\bibitem[\protect\citeauthoryear{Boyd}{2000}]{REFFBoyd}
 Boyd I.D.~ 2000, Earth Moon Planets, 82/83, 93

\bibitem[\protect\citeauthoryear{Dimant et al.}{2009}]{REFFDimant}
 Dimant Y.S., Oppenheim M.M., Milikh, G.M.\ 2009, Ann. Geophys., 27, 279

\bibitem[\protect\citeauthoryear{Dokuchayev}{1960}]{REFFDokuchayev}
 Dokuchayev V.P.\ 1960, Dokl. Akad. Nauk SSSR, 131, 78

\bibitem[\protect\citeauthoryear{Farbar \& Boyd}{2010}]{REFFFarbar}
  Farbar E.D., Boyd I.D.\ 2010,  Phys. Fluids, 22, 106101

\bibitem[\protect\citeauthoryear{Forrest et al.}{1992}]{REFForrest}
 Forrest R.G., Russ R.L., Patrick J.E.\ 1992,  J. Phys. Chem. Ref. Data, 21, 1005

\bibitem[\protect\citeauthoryear{Gao \& Mathews}{2015}]{REFFGao}
 Gao B., Mathews J.D.\ 2015,  MNRAS,  446, 3404

\bibitem[\protect\citeauthoryear{Hedin}{1991}]{REFFHedin}
 Hedin A.E.\ 1991, J. Geophys. Res., 96, 1159

\bibitem[\protect\citeauthoryear{Itikawa \& Ichimura}{1990}]{REFItikawa}
 Itikawa Y., Ichimura A.\  1990, J. Phys. Chem. Ref. Data, 19, 637

\bibitem[\protect\citeauthoryear{Jehin et al.}{2007}]{REFFJehin}
 Jehin E., et al.\ 2007, Adv. Space Res., 39, 550

\bibitem[\protect\citeauthoryear{Jenniskens}{2002}]{Jenniskens2002}
 Jenniskens P.\ 2002,  WGN, 30, 218

\bibitem[\protect\citeauthoryear{Jenniskens}{2004}]{REFFJenniskensD}
 Jenniskens P.\ 2004,  Adv. Space Res., 33, 1444

\bibitem[\protect\citeauthoryear{Jenniskens et al.}{2004}]{REFFJenniskensB}
 Jenniskens P., Laux C.O., Wilson M.A., Schaller E.L.~ 2004, Astrobiology, 4, 81

\bibitem[\protect\citeauthoryear{Jenniskens \& Mandell}{2004}]{REFFJenniskensC}
 Jenniskens P., Mandell A.M.\ 2004, Astrobiology, 4, 123

\bibitem[\protect\citeauthoryear{Kasuga et al.}{2007}]{REFKasuga}
 Kasuga T., Iijima T., Watanabe J.\ 2007, Astron. Astrophys., 474, 639
 
\bibitem[\protect\citeauthoryear{Kelley and Price}{2017}]{Kelley}
 Kelley M.~C., Price C.\ 2017, Geophys. Res. Lett., 44, 2987  

\bibitem[\protect\citeauthoryear{Kero et al.}{2008}]{REFFKero}
  Kero J., et al.\ 2008,  Geophys. Res. Lett., 35, L04101

\bibitem[\protect\citeauthoryear{Lamar \& Romig}{1964}]{REFRAND}
 Lamar D.L, Romig M.F.\ 1964, Meteoritics, 2, 127

\bibitem[\protect\citeauthoryear{LeBlanc et al.}{2000}]{REFFLeBlanc}
  LeBlanc A.G., et al.\ 2000,  MNRAS, 313, L9

\bibitem[\protect\citeauthoryear{Marshall \& Close}{2015}]{REFFMarshall}
 Marshall R.A., Close S.\ 2015, J. Geophys. Res., 120, 5931

\bibitem[\protect\citeauthoryear{Marshall et al.}{2017}]{REFFClose}
 Marshall R.~A., Brown P., Close S.\ 2017, Planet. Space Sci., 143, 203

\bibitem[\protect\citeauthoryear{Mendis et al.}{2005}]{REFFMendis}
 Mendis D.A., Wong W.-H., Rosenberg M., Sorasio G.\ 2005, J. Atmos. Solar Terr. Phys., 67, 1178

\bibitem[\protect\citeauthoryear{Oppenheim and Dimant}{2015}]{Oppenheim}
 Oppenheim M.M., Dimant Y.S.\ 2015, Geophys. Res. Lett., 42, 681

\bibitem[\protect\citeauthoryear{Pilyugin \& Baulin}{1993}]{PilyuginBaulin}
 Pilyugin N. N., Baulin N. N.\ 1993, Solar System Research, 27, 558
 
\bibitem[\protect\citeauthoryear{Pilyugin}{1994}]{Pilyugin93}
 Pilyugin N.N.\ 1994, High Temperature, 32, 110

\bibitem[\protect\citeauthoryear{Plane}{2012}]{REFFPlane}
 Plane J.M.C.\ 2012, Chem. Soc. Rev., 41, 6507

\bibitem[\protect\citeauthoryear{Popova et al.}{2013}]{REFFPopova}
 Popova O.P., et al.\ 2013, Science, 342, 1069

\bibitem[\protect\citeauthoryear{Popova et al.}{2000}]{REFFPopovaVapor}
 Popova O.P., Sidneva S.N., Shuvalov V.V., Strelkov A.S.\ 2000, Earth Moon Planets, 82/83, 109

\bibitem[\protect\citeauthoryear{Popova et al.}{2007}]{REFFPopovaSputter}
 Popova O.O., Strelkov A.S. Sidneva S.N.\ 2007, Adv. Space Res., 39, 567

\bibitem[\protect\citeauthoryear{Qin et al.}{2014}]{REFFQin}
  Qin J., Pasko V.P., McHarg M.G., Stenbaek-Nielsen H.C.\ 2014, Nature Comm., 5, id.3740

\bibitem[\protect\citeauthoryear{Rees}{1989}]{REFFRees}
 Rees M.H.\ 1989, Physics and chemistry of the upper atmosphere,  Cambridge University Press

\bibitem[\protect\citeauthoryear{Schunk \& Nagy}{2000}]{REFFSchunk}
  Schunk R.W., Nagy A.F.\  2000, Ionospheres: Physics, Plasma Physics, and Chemistry, Cambridge University Press

\bibitem[\protect\citeauthoryear{Serov \& Yavor}{1991}]{SerovYarov91}
 Serov Yu. L., Yavor I.P.\ 1991,  Soviet Physics. Technical physics., 36, 972

\bibitem[\protect\citeauthoryear{Serov \& Yavor}{1996}]{SerovYarov96}
 Serov Yu. L., Yarov M. P.\ 1996, Technical Physics, 41, 1183

\bibitem[\protect\citeauthoryear{Solomon}{2001}]{REFFSolomon}
 Solomon S.C.\ 2001, J. Geophys. Res., 106, 107

\bibitem[\protect\citeauthoryear{Spurn\'{y} et al.}{2000}]{REFFSpurny2000}
 Spurn\'{y} P., et al.\ 2000, Meteorit. Planet. Sci., 35, 1109

\bibitem[\protect\citeauthoryear{Spurn\'{y} \& Ceplecha}{2008}]{REFFSpurny}
  Spurn\'{y} P. Ceplecha Z.\ 2008,  Astron. Astrophys., 489, 449

\bibitem[\protect\citeauthoryear{Stenbaek-Nielsen and Jenniskens}{2004}]{REFFStenbaekNielsen}
 Stenbaek-Nielsen H.C., Jenniskens P.A.\ 2004, Adv. Space Res., 33, 1459

\bibitem[\protect\citeauthoryear{Stokan \& Campbell-Brown}{2014}]{REFFStokan}
  Stokan E., Campbell-Brown M.D.\ 2014,  Icarus, 232, 1

\bibitem[\protect\citeauthoryear{Surzhikov}{2014}]{REFFSurzhikov}
 Surzhikov S.T.\ 2014, 44th AIAA Fluid Dynamics Conference, AIAA Aviation (AIAA 2014-2636)

\bibitem[\protect\citeauthoryear{ter Haseborg \& Trinks}{1980a}]{Haseborg1}
 ter Haseborg J. L., Trinks H.\ 1980, IEEE Transactions on Aerospace and Electronic Systems, AES-16, 227

\bibitem[\protect\citeauthoryear{ter Haseborg \& Trinks}{1980b}]{Haseborg2}
 ter Haseborg J. L., Trinks H.\ 1980, IEEE Transactions on Aerospace and Electronic Systems, AES-16, 750

\bibitem[\protect\citeauthoryear{Th{\'e}bault et al.}{2015}]{REFFigrf} 
 Th{\'e}bault E., et al.\ 2015, Earth, Planets, and Space, 67, id.\#79

\bibitem[\protect\citeauthoryear{Vinkovi\'{c}}{2007}]{REFFVinkovicA}
  Vinkovi\'{c} D.\ 2007,  Adv. Space Res., 39, 574

\bibitem[\protect\citeauthoryear{Wiesemann}{2014}]{Wiesemann}
Wiesemann K.\ 2014, CERN Yellow Report CERN-2013-007, 85; arXiv:1404.0509

\bibitem[\protect\citeauthoryear{Zgrabli\'{c} et al.}{2002}]{REFFZgrablic}
 Zgrabli\'{c}, G., et al.\ 2002, J. Geophys. Res., 107, 1124


\end{thebibliography}
\end{document}